\documentclass[12pt]{article}
\usepackage{epsfig,amsfonts,amsmath,array,mathrsfs}
\newcommand{\pbs}[1]{\let\temp=\\#1\let\\=\temp}
\renewcommand{\theequation}{\thesection.\arabic{equation}}
\def\be{\begin{equation}}\def\ee{\end{equation}}
\def\cvp{\raise 2pt\hbox{,}}

 \def\Tr{\mathop{\rm Tr}\nolimits}
 
 \def\im{\mathop{\rm Im}\nolimits}
\def\re{\mathop{\rm Re}\nolimits} 
\def\diag{\mathop{\rm diag}\nolimits}

\def\d{{\rm d}}
\def\nn{{\cal N}} 

\def\bigllangle{\bigl\langle\!\bigl\langle}
\def\bigrrangle{\bigr\rangle\!\bigr\rangle}
\def\Bigllangle{\Bigl\langle\!\!\Bigl\langle}
\def\Bigrrangle{\Bigr\rangle\!\!\Bigr\rangle} 
\def\Nf{N_{\mathrm f}}
 \def\uN{{\rm U}(N)} 
  
\def\la{\lambda}
\def\mac{\text{mac}}\def\mic{\text{mic}}
\def\Zmic{Z_{\text{mic}}}\def\Zmac{Z_{\text{mac}}} \def\Fmic{{\mathcal
F}_{\text{mic}}}\def\Fmac{{\mathcal F}_{\text{mac}}}
\def\Rmic{R_{\text{mic}}}\def\Rmac{R_{\text{mac}}}
\def\Smic{S_{\text{mic}}}\def\Smac{S_{\text{mac}}}
\def\lmic{\lambda_{\text{mic}}}\def\lmac{\lambda_{\text{mac}}}
\def\Lmic{\mathscr L_{\text{mic}}}\def\Lmac{\mathscr L_{\text{mac}}}
\def\wmic{W_{\text{mic}}}
\def\wmac{W_{\text{mac}}}

\def\a{\boldsymbol{a}}
\def\s{\boldsymbol{s}}\def\g{\boldsymbol{g}}
 \def\t{\boldsymbol{t}}
\def\uR{\text{U}(1)_{\text R}}\def\uA{\text{U}(1)_{\text A}}
\def\u{\text{U}(1)}
 \def\eps{\epsilon}\def\veps{\varepsilon}

\def\vevab#1{\bigl\langle\a\big|#1\big|\a\bigr\rangle}

\def\vevsb#1{\bigllangle\s\big|#1\big|\s\bigrrangle}

\def\vevabe#1{\bigl\langle\a\big|#1\big|\a\bigr\rangle_{\epsilon}}

\def\vevsbe#1{\bigllangle\s\big|#1\big|\s\bigrrangle_{\varepsilon}}
\def\vevsBe#1{\Bigllangle\s\Big|#1\Big|\s\Bigrrangle_{\varepsilon}}
\def\vevaBe#1{\Bigl\langle\a\Big|#1\Big|\a\Bigr\rangle_{\epsilon}}
\def\cpart{\vec{\mathsf p}}
\def\plb#1#2#3{{\it Phys.\ Lett.\ }{\bf B #1} (#2) #3}
\def\npb#1#2#3{{\it Nucl.\ Phys.\ }{\bf B #1} (#2) #3}

\def\jhep#1#2#3{{\it J. High Energy Phys.\ }{\bf #1} (#2) #3}

\def\atmp#1#2#3{{\it Adv.\ Theor.\ Math.\ Phys.\ }{\bf #1} (#2) #3}
\def\cmp#1#2#3{{\it Comm.\ Math.\ Phys.\ }{\bf #1} (#2) #3}

\def\nca#1#2#3{{\it Nuovo Cim.\ }{\bf #1} (#2) #3}

\begin{document}
%
%\pdfgraphics 
%\pagenumbering{roman}
%
\pagestyle{empty}
{\parskip 0in

\hfill LPTENS-07/35

\hfill arXiv:0709.0472 [hep-th]}

\vfill
\begin{center}
%{\sffamily\large\bfseries SPECTRAL ASYMMETRY AND SUPERSYMMETRY}
{\LARGE Extended $\nn=1$ super Yang-Mills theory}

%\medskip

%{\LARGE in $\nn=1$ gauge theories}

\vspace{0.4in}

Frank \textsc{Ferrari}%{\renewcommand{\thefootnote}{$\!\!\dagger$}
%\footnote{On leave of absence from Centre National de la Recherche
%Scientifique, Laboratoire de Physique Th\'eorique de l'\'Ecole Normale
%Sup\'erieure, Paris, France.}} 
\\
\medskip
{\it Service de Physique Th\'eorique et Math\'ematique\\
Universit\'e Libre de Bruxelles and International Solvay Institutes\\
Campus de la Plaine, CP 231, B-1050 Bruxelles, Belgique
%{\renewcommand{\thefootnote}{$\!\!\!\!\!\dagger$}
%\footnote{New permanent address}
}\\
%\smallskip
%School of Natural Sciences, Institute for Advanced Study\\
%Einstein Drive, Princeton, New Jersey 08540, USA}\\
\smallskip
{\tt frank.ferrari@ulb.ac.be}
\end{center}
\vfill\noindent

We solve a generalization of ordinary $\nn=1$ super Yang-Mills theory
with gauge group $\uN$ and an adjoint chiral multiplet $X$ for which
we turn on both an arbitrary tree-level superpotential term
$\int\!\d^{2}\theta\Tr W(X)$ and an arbitrary field-dependent gauge
kinetic term $\int\!\d^{2}\theta\Tr V(X)W^{\alpha}W_{\alpha}$. When
$W=0$, the model reduces to the extended Seiberg-Witten theory
recently studied by Marshakov and Nekrasov. We use two different
points of view: a ``macroscopic'' approach, using generalized anomaly
equations, the Dijkgraaf-Vafa matrix model and the glueball
superpotential; and the recently proposed ``microscopic'' approach,
using Nekrasov's sum over colored partitions and the quantum
microscopic superpotential. The two formalisms are based on completely
different sets of variables and statistical ensembles. Yet it is shown
that they yield precisely the same gauge theory correlators. This
beautiful mathematical equivalence is a facet of the open/closed
string duality. A full microscopic derivation of the non-perturbative
$\nn=1$ gauge dynamics follows.

\vfill

%\noindent Gravit\'e, Th\'eories de Jauge et Cordes,} 
%Les Houches summer school 2001, Session LXXVI.
\medskip
%
%\vfill
\begin{flushleft}
\today
\end{flushleft}
\newpage\pagestyle{plain}
\baselineskip 16pt
\setcounter{footnote}{0}

\section{General presentation}
\setcounter{equation}{0}

In two recent papers \cite{mic1,mic2}, the author and collaborators
have proposed a general microscopic approach to the solution of
$\nn=1$ super Yang-Mills theories. In \cite{mic1}, it was explained
how to apply Nekrasov's instanton technology \cite{nekrasova} to
$\nn=1$, including in the strongly coupled vacua. The formalism is
based on a microscopic quantum superpotential $\wmic$ whose saddle
points are in one-to-one correspondence with the full set of quantum
vacua of the theory. In \cite{mic2}, explicit calculations were made
up to two instantons, and it was shown that the results agree to this
order with the predictions made using a totally different formalism
based on the Dijkgraaf-Vafa matrix model and glueball superpotential
\cite{DV,CDSW}. The main purpose of the present work is to present a
proof of the exact equivalence between the two formalisms. This yields
a full microscopic derivation of the exact results in $\nn=1$ gauge
theories, including a non-perturbative justification of the
Dijkgraaf-Vafa matrix model, the generalized anomaly equations and the
Dijkgraaf-Vafa glueball superpotential.

\subsection{The model}

We shall focus, as in \cite{mic1,mic2}, on the $\nn=1$ theory with
gauge group $\uN$ and one adjoint chiral superfield $X$. The basic
chiral operators are given by \cite{CDSW}\footnote{One can also
introduce the chiral operators $\Tr W^{\alpha}X^{k}$, but they have
zero expectation values due to Lorentz invariance. Non-trivial
expectation values could be obtained by turning on Lorentz violating
couplings $t^{\alpha}_{k}\Tr W_{\alpha}X^{k}$ in the tree-level
superpotential. The resulting model can be studied straightforwardly
using our methods, but we shall not do it here for the sake of
simplicity.}
\begin{align}\label{defu} u_{k} &= \Tr X^{k}\, ,\\\label{defv}
v_{k} & = -\frac{1}{16\pi^{2}}\Tr W^{\alpha}W_{\alpha}X^{k}\,
,\end{align}
where $W^{\alpha}$ is the vector chiral superfield which contains the
gauge field and whose lowest component is the gluino field. Our main
goal is to compute the gauge theory expectation values of the above
operators, that are conveniently encoded in the generating functions
\begin{align}\label{Rdef} R(z) &= \sum_{k\geq 0}\frac{\langle
u_{k}\rangle}{z^{k+1}}\,\cvp\\ \label{Sdef} S(z) &= \sum_{k\geq
0}\frac{\langle v_{k}\rangle}{z^{k+1}}\,\cdotp\end{align}
The theory is usually studied with an arbitrary tree-level
superpotential $\Tr W(X)$ for the field $X$, which amounts to
introducing arbitrary couplings to the scalar operators \eqref{defu}.
For our purposes, it is extremely natural to introduce arbitrary
couplings to the generalized glueball operators \eqref{defv} as well.
The tree-level lagrangian that we consider is thus of the form
\be\label{Lag1} L =
\frac{1}{4\pi}\im\int\!\d^{2}\theta\,\Tr\bigl(
\tau(X)W^{\alpha}W_{\alpha}\bigr)
+ 2N\re\int\!\d^{2}\theta\, \Tr W(X)\, ,\ee
where $\tau(X)$ and $W(X)$ are arbitrary polynomials in $X$. Instead
of the field-dependent coupling $\tau$, it is convenient to work with 
the polynomial $V$ defined by
\be\label{Vdef} NV(z) = 2i\pi\tau(z)\, .\ee
The lagrangian can then be written as
\be\label{Lag2} L = 2N\re\int\!\d^{2}\theta\,\mathscr W\ee
with
\be\label{treeWgen}\mathscr W = - \frac{1}{16\pi^{2}}\Tr V(X)
W^{\alpha}W_{\alpha}+\Tr W(X) \, .\ee
Using the parametrization
\begin{align}\label{Vexp}V(z) &= \la_{-1} +
\sum_{k=0}^{d_{V}}\frac{\la_{k}}{k+1}\, z^{k+1}\, ,\\ \label{Wexp}
W(z) &= \sum_{k=0}^{d_{W}}\frac{g_{k}}{k+1}\, z^{k+1}\, ,\end{align}
where the degrees $d_{V}$ and $d_{W}$ can be arbitrary, the
superpotential \eqref{treeWgen} takes the form
\be\label{treeWuv} \mathscr W = \la_{-1}v_{0} +\sum_{k\geq
0}\frac{\la_{k}}{k+1}\, v_{k+1}+ \sum_{k\geq 0}\frac{g_{k}}{k+1}\,
u_{k+1}\, . \ee
It is also useful to introduce a polynomial
\be\label{text} t(z) = \sum_{k\geq 1}\frac{t_{k}}{k+1}\, z^{k+1}\ee
satisfying the relation
\be\label{tVrel} t''(z) = NV(z) = 2i\pi\tau(z)\, .\ee
We shall use $t(z)$ or $V(z)$ interchangeably according to
convenience, with
\be\label{tlambdarel} t_{1}=N\la_{-1}\, ,\quad t_{k} =
\frac{N\la_{k-2}}{k(k-1)}\quad \text{for}\ k\geq 2\, .\ee
Let us note that the usual instanton factor is given by
\be\label{qdef} q = e^{t_{1}} = e^{N\la_{-1}}\, .\ee
The model \eqref{Lag1} has useful $\uA$ and $\uR$ global symmetries.
The charges of the superspace coordinates $\theta^{\alpha}$, of the
various fields and couplings, and of any superpotential $w$ one may
wish to consider are given in the following table,
\be\label{asign}
\begin{matrix}
& \theta & W^{\alpha} & X & u_{k} & v_{k} & g_{k} & \la_{k},\ k\geq 0 
& q & w\\
{\rm U}(1)_{\rm A} & 0 & 0 & 1 & k & k & -k-1 & -k-1 & 2N & 0 \\
{\rm U}(1)_{\rm R} & 1 & 1 & 0 & 0 & 2 & 2 & 0 & 0 &\hphantom{,\,} 2
\, . \end{matrix}\ee

When $W=0$, the theory \eqref{Lag1} has $\nn=2$ supersymmetry, with
tree-level prepotential
\be\label{Ftree}\mathcal F_{\nn=2} (X) = t(X)\, .\ee
This ``extended'' Seiberg-Witten theory was studied recently from the
microscopic point of view by Marshakov and Nekrasov in \cite{MN}, and
their results will be particularly useful for us. When $W$ is turned
on, the $\nn=2$ moduli space is lifted, and the theory has a discrete
set of vacua. Classically, the vacua are labeled by the numbers
$N_{i}$ of eigenvalues of $X$ that sit at given critical points of
$W$. In such vacua, the gauge group is broken from $\uN$ down to
$\text U(N_{1})\times\cdots\times\text{U}(N_{d_{W}})$. The \emph{rank}
of the vacuum is defined to be the number of non-zero integers
$N_{i}$. Since a mass gap is created in each non-abelian unbroken
factor, it coincides with the rank of the low energy gauge group
$\u^{r}$. Moreover, chiral symmetry breaking generates an $N_{i}$-fold
degeneracy for each $\text{U}(N_{i})$ factor. This will be explicitly
demonstrated later. The quantum vacua corresponding to the integers
$N_{i}$ are thus labeled as $|N_{i},k_{i}\rangle$ with $0\leq
k_{i}\leq N_{i}-1$. Moreover, when $d_{V}\geq 1$, we may find that new
vacua appear at the quantum level. These vacua go to infinity in field
space in the classical limit.

The solution of the model \eqref{Lag1} can be found using two a priori
completely different approaches. One approach is motivated by the
closed string dual of the gauge theory and is natural from the
long-distance, macroscopic point of view. It is based on the
Dijkgraaf-Vafa matrix model and the use of the glueball superpotential
\cite{DV}, or equivalently on the geometric transition picture and the
flux superpotential in the dual closed string background \cite{CIV}.
We call this approach the \emph{macroscopic formalism}. It is very
difficult to justify this formalism from first principles. The second
approach amounts to computing directly the relevant gauge theory path
integrals. It is based on Nekrasov's sums over colored partitions
\cite{nekrasova} and the microscopic quantum superpotential
\cite{mic1}. This is natural from the short-distance point of view and
thus we call this approach the \emph{microscopic formalism}. The
microscopic formalism provides rigorous, first-principle derivations
of the non-perturbative gauge theory dynamics.

The goal of the present paper is to prove the equivalence between the
two formalisms. Since the microscopic and macroscopic set-ups are
equivalent to the open and closed string descriptions respectively,
the mathematical equivalence we are going to derive is a beautiful
facet of the open/closed string duality, in a rare case where a
complete understanding can be achieved.

Let us now present briefly the main ingredients of the two formalisms.
Full details and justifications will be given in later Sections.

\noindent\textsc{Notation}: In the following, when we have an indexed
family of parameters, we use a non-indexed boldface letter to
represent all the parameters at once. For example $\g$ denotes all the
$g_{k}$, and $\t$ all the $t_{k}$.

\subsection{The macroscopic formalism}
\label{macfor}

In the macroscopic formalism, the basic, natural variables are the
generalized glueball operators \eqref{defv}. Their expectation values
for fixed gluino condensates $s_{i}$ in the unbroken factors of the
gauge group are given in terms of averages over the statistical
ensemble of a random hermitian matrix $M$ of size $n\times n$ as
\begin{align}\label{vkmac}
&v_{k,\,\mac}(\s,\g,\veps)=N\veps\,\vevsbe{\Tr X^{k}} =
\frac{N\veps}{\Zmac}\int\!\d\mu_{\mac}^{M}\Tr M^{k}\, , \\ \label{Zmac}
&\Zmac(\s,\g,\veps) = \int\!\d\mu_{\mac}^{M} =
\exp\frac{\Fmac(\s,\g,\veps)}{\veps^{2}}\, \cdotp\end{align}
The ``macroscopic'' measure is given in terms of the components of the
matrix $M$ and the tree-level superpotential $W$,
\be\label{macmeadef}\d\mu_{\mac}^{M} = \prod_{I=1}^{n}\d
M_{II}\prod_{1\leq I<J\leq n}\d\!\re M_{IJ}\,\d\!\im M_{IJ}\,
\exp\Bigl(-\frac{1}{\veps}\Tr W(M)\Bigr)\, .\ee
The parameter $\veps$, which can be interpreted as the strength of
some particular supergravity background \cite{DS}, is related to the
size of the matrix $M$,
\be\label{vepsnrel} \veps = \frac{s}{n}\,\cvp\ee
with
\be\label{ssirel} s = \sum_{i}s_{i}\, .\ee
The precise prescription to compute \eqref{vkmac} is as follows. If
one wishes to describe a rank $r$ vacuum, then one must expand the
matrix integral around the corresponding classical saddle point
$|N_{1},\ldots,N_{r}\rangle$, by putting $n_{i}=s_{i}/\veps$
eigenvalues of the matrix $M$ at the critical point of $W$
corresponding to the integer $N_{i}$. One then considers the large
$n$, or small $\veps$, 't~Hooft's genus expansion.

From \eqref{vkmac}, \eqref{Zmac} and \eqref{macmeadef}, we obtain
immediately relations valid for any $\veps$,
\be\label{vkFmac} v_{k,\,\mac}(\s,\g,\veps) =
-Nk\frac{\partial\Fmac}{\partial g_{k-1}}\,\cvp\quad k\geq 1\, .\ee
These relations are the macroscopic analogue of the Matone's relations
\cite{matone}, see \eqref{ukFmic} and below.

From \eqref{vkmac}, we can get the generating function
\be\label{Smacdef} \Smac(z;\s,\g,\veps) = \sum_{k\geq
0}\frac{v_{k,\,\mac}(\s,\g,\veps)}{z^{k+1}}\,\cdotp\ee
Most relevant to us will be the planar $\veps\rightarrow 0$ limit
\begin{align}\label{vkgt} v_{k,\,\mac}(\s,\g) &=
\lim_{\veps\rightarrow 0}v_{k,\,\mac}(\s,\g,\veps)\, ,\\
\label{Smacgt} \Smac(z;\s,\g) & = \lim_{\veps\rightarrow
0}\Smac(z;\s,\g,\veps)\, .\end{align}
Note the following important feature: the function $\Smac$ does
\emph{not} depend on the parameters $\t$ that enter the tree-level
lagrangian \eqref{Lag1}.

The next step is to introduce the macroscopic quantum superpotential
$\wmac$, which is nothing but the glueball superpotential.
In terms of the ``macroscopic'' one-form
\be\label{macformdef} \la_{\mac} = \Smac(z;\s,\g)\,\d z\, ,\ee
it is given by
\be\label{wmacdef} \wmac(\s,\g,\t) =
\frac{1}{2i\pi}\oint_{\alpha}V\lmac -
\sum_{i}N_{i}\frac{\partial\Fmac}{\partial s_{i}}\,\cvp\ee
where $\alpha$ will always denote a large contour at infinity in the
$z$-plane. The expectation values of the operators \eqref{defu} are
given by
\be\label{ukmac} u_{k,\,\mac}(\s,\g,\t) =
k\frac{\partial\wmac}{g_{k-1}}\, \cvp \ee
with associated generating function
\be\label{Rmacdef}\Rmac(z;\s,\g,\t) = \sum_{k\geq
0}\frac{u_{k,\,\mac}(\s,\g,\t)}{z^{k+1}}\,\cdotp\ee
Unlike $\Smac$, $\Rmac$ does depend, linearly, on the parameters $\t$.

The parameters $\s$ are determined by solving the equations
\be\label{macqem} \frac{\partial\wmac}{\partial
s_{i}}\bigl(\s=\s^{*}\bigr) = 0\, .\ee
These equations have in general several solutions, that are in
one-to-one correspondence with the quantum vacua of fixed rank $r$.
The on-shell generating functions
\begin{align}\label{onshellmacS} \Smac^{*}(z;\g,\t) &=
\Smac(z;\s^{*},\g)\, ,\\\label{onshellmacR} \Rmac^{*}(z;\g,\t) &=
\Rmac(z;\s^{*},\g,\t)\, ,\end{align}
are conjectured to coincide with the corresponding gauge theory
observables,
\begin{align}\label{conjmacS} S(z;\g,\t)  &= \Smac^{*}(z;\g,\t)\, ,\\
\label{conjmacR} R(z;\g,\t)  &= \Rmac^{*}(z;\g,\t)\, .\end{align}
Of course both $\Smac^{*}$ and $\Rmac^{*}$ depend non-linearly on $\t$
because $\s^{*}$ gets a non-trivial $\t$-dependence upon solving
\eqref{macqem}.

\subsection{The microscopic formalism}
\label{micfor}

In the microscopic formalism, the basic, natural variables are the
operators \eqref{defu}. Their expectation values for fixed boundary
conditions at infinity $a_{i}$ for the field $X$,\footnote{We work in 
euclidean space-time.}
\be\label{bcdef} X_{\infty} = \diag (a_{1},\ldots,a_{N}) = \diag\a\,
,\ee
are given in terms of averages over the statistical ensemble of random
colored partitions endowed with a suitable generalized Plancherel
measure as
\begin{align}\label{ukmic} &u_{k,\,\mic}(\a,\t,\eps)=\vevabe{\Tr
X^{k}} = \frac{1}{\Zmic}\sum_{\cpart}\mu_{\mic}^{\cpart}\,
u_{k,\,\cpart}\, , \\ \label{Zmic} &\Zmic(\a,\t,\eps)
=\sum_{\cpart}\mu_{\mic}^{\cpart} =
\exp\frac{\Fmic(\a,\t,\eps)}{\eps^{2}}\, \cdotp\end{align}
A colored partition $\cpart$ is a collection of $N$ ordinary
partitions $\mathsf p_{i}$, $\cpart=(\mathsf p_{1},\ldots,\mathsf
p_{N})$, which are characterized by integers $p_{i,\alpha}$ satisfying
\begin{gather}\label{pialpdef} p_{i,1}\geq p_{i,2}\geq\cdots\geq p_{i,\tilde
p_{i,1}}>p_{i,\tilde p_{i,1}+1}=0\, ,\\
\sum_{\alpha=1}^{\tilde p_{i,1}} p_{i,\alpha} = |\mathsf p_{i}|\,
.\end{gather}
The integer $|\mathsf p_{i}|$ is called the size of the partition
$\mathsf p_{i}$, and
\be\label{sizep} |\cpart| = \sum_{i=1}^{N}|\mathsf p_{i}|\ee
is the size of the colored partition $\cpart$. To each partition
$\mathsf p_{i}$, it is convenient to associate a Young tableau
$Y_{\mathsf p_{i}}$ with $p_{i,\alpha}$ boxes in the row number
$\alpha$ (the uppermost row being the last, shortest row). The number
of boxes in the column number $\beta$ is then denoted by $\tilde
p_{i,\beta}$ (the rightmost column being the last, shortest column).
The integers $\tilde p_{i,\beta}$ automatically satisfy
\begin{gather}\label{pibetdef} \tilde p_{i,1}\geq \tilde
p_{i,2}\geq\cdots\geq \tilde p_{i,p_{i,1}}>\tilde p_{i,p_{i,1}+1}=0\,
,\\ \label{sizepi} \sum_{\beta=1}^{p_{i,1}} \tilde p_{i,\beta} =
|\mathsf p_{i}|\, .\end{gather}
In \eqref{ukmic} and \eqref{Zmic}, the ``microscopic'' measure is
given in terms of the integers characterizing the colored partition
$\cpart$ and the tree-level gauge kinetic term. Explicitly, we have
\be\label{munurel} \mu_{\mic}^{\cpart} =
\bigl(\nu_{\mic}^{\cpart}\bigr)^{2}\ee
with
\begin{multline}\label{micmeadef} \nu_{\mic}^{\cpart}
=\frac{1}{\eps^{|\cpart|}}
\prod_{i=1}^{N}\Biggl[\prod_{\Box_{(\alpha,\beta)}\in Y_{\mathsf
p_{i}}}\frac{1}{p_{i,\alpha}-\beta+\tilde p_{i,\beta}-\alpha+1}
\prod_{j\not = i}\frac{1}{a_{i}-a_{j}+\eps(\beta-\alpha)}\Biggr] \times
\\
\prod_{i<j}\prod_{\alpha=1}^{\tilde p_{i,1}}\prod_{\beta=1}^{p_{j,1}}
\frac{\bigl( a_{i}-a_{j}+\eps (\tilde p_{j,\beta}-\alpha-\beta
+1)\bigr)\bigl(a_{i}-a_{j}+\eps(p_{i,\alpha}-\beta-\alpha+1)\bigr)}
{\bigl(a_{i}-a_{j}+\eps(1-\alpha-\beta)\bigr)
\bigl(a_{i}-a_{j}+\eps(\tilde p_{j,\beta}-\alpha+ p_{i,\alpha}-\beta
+1)\bigr)}\,\times\\
\exp\Bigl(\frac{1}{2\eps^{2}}\sum_{k\geq 1}
\frac{t_{k}}{k+1}\, u_{k+1,\,\cpart} \Bigr)
\end{multline}
and
\begin{multline}\label{ukpart} u_{k,\,\cpart} = \sum_{i=1}^{N}\biggl[
a_{i}^{k} + \sum_{\alpha=1}^{\tilde p_{i,1}}\Bigl(\bigl( a_{i} +
\eps(p_{i,\alpha}-\alpha+1)\bigr)^{k} - \bigl(a_{i}+\eps
(p_{i,\alpha}-\alpha)\bigr)^{k}\\ + \bigl( a_{i} - \eps\alpha\bigr)^{k}
- \bigl( a_{i}-\eps(\alpha-1)\bigr)^{k}\Bigr)\biggr]\, .\end{multline}
The parameter $\eps$ can be interpreted as being the strength of some
particular supergravity background, the so-called $\Omega$-background
\cite{nekrasova}. It is different from the supergravity background
governed by the parameter $\veps$ in the macroscopic formalism. In
particular, $\veps$ is associated with a non-trivial space-time
curvature whereas $\eps$ is not.

Let us note that
\be\label{u2parteq} u_{2,\,\cpart} = \sum_{i=1}^{N}a_{i}^{2} +
2\eps^{2}|\cpart|\, .\ee
Using \eqref{micmeadef}, \eqref{munurel} and \eqref{qdef}, this
implies that the dependence in the instanton factor in the sums
\eqref{ukmic} and \eqref{Zmic} is given by $q^{|\cpart|}$. Thus
colored partitions of size $k$ contribute to the $k^{\text{th}}$
instanton order.

From \eqref{ukmic}, \eqref{Zmic} and \eqref{micmeadef}, we obtain
immediately a set of generalized Matone's relations \cite{matone},
valid for any $\eps$,
\be\label{ukFmic} u_{k,\,\mic}(\a,\t,\eps) =
2k\frac{\partial\Fmic}{\partial t_{k-1}}\,\cvp\quad k\geq 2\, .\ee
The usual Matone's relation corresponds to $k=2$, $t_{k'}=0$ for
$k'\geq 2$ and $\eps=0$. It was shown to be valid at finite $\eps$ in 
\cite{fucitomatone}.

From \eqref{ukmic} we can get the generating function
\be\label{Rmicdef}\Rmic(z;\a,\t,\eps) = \sum_{k\geq
0}\frac{u_{k,\,\mic}(\a,\t,\eps)}{z^{k+1}}\,\cdotp\ee
Most relevant to us will be the limit $\eps\rightarrow 0$ of vanishing
$\Omega$-background
\begin{align}\label{ukgt} u_{k,\,\mic}(\a,\t) &=\vevab{\Tr X^{k}}=
\lim_{\eps\rightarrow 0}u_{k,\,\mic}(\a,\t,\eps)\, ,\\
\label{Rmicgt} \Rmic(z;\a,\t) & = \lim_{\eps\rightarrow
0}\Rmic(z;\a,\t,\eps)\, .\end{align}
Note the following important feature: the function $\Rmic$ does
\emph{not} depend on the parameters $\g$ that enter the tree-level
lagrangian \eqref{Lag1}.

The next step is to introduce the microscopic
quantum superpotential $\wmic$ \cite{mic1}. In terms of the
``microscopic'' one-form
\be\label{micformdef} \lmic = z\Rmic(z;\a,\t)\,\d z\, ,\ee
it is given by
\be\label{wmicdef} \wmic(\a,\g,\t) = \vevab{\Tr W(X)}
=\frac{1}{2i\pi}\oint_{\alpha}\frac{W\lmic}{z}\, \cdotp\ee
The expectation values of the operators \eqref{defv} are given by
\be\label{vkmic}\begin{split}v_{0,\,\mic}(\a,\g,\t) &=
\frac{\partial\wmic}{\partial\la_{-1}}\,\cvp\\
v_{k,\,\mic}(\a,\g,\t) &
=k\frac{\partial\wmic}{\partial\la_{k-1}}\quad \text{for}\ k\geq 1\,
,\end{split} \ee
or equivalently by
\be\label{vkmicbis} v_{k,\,\mic}(\a,\g,\t) = \frac{N}{k+1}
\frac{\partial\wmic}{\partial t_{k+1}}\, \cdotp\ee
The associated generating function is
\be\label{Smicdef}\Smic(z;\a,\g,\t) = \sum_{k\geq
0}\frac{v_{k,\,\mic}(\a,\g,\t)}{z^{k+1}}\,\cdotp\ee
Unlike $\Rmic$, $\Smic$ does depend, linearly, on the parameters $\g$.

The parameters $\a$ are determined by solving the equations
\be\label{micqem} \frac{\partial\wmic}{\partial
a_{i}}\bigl(\a=\a^{*}\bigr) = 0\, .\ee
These equations have in general several solutions, that are in
one-to-one correspondence with the full set of quantum vacua of the
theory \cite{mic1}. This is in sharp contrast with the equations
\eqref{macqem}, that yield the vacua for fixed values of the rank $r$
only. The on-shell generating functions
\begin{align}\label{onshellmicR} \Rmic^{*}(z;\g,\t) &=
\Rmic(z;\a^{*},\t)\, ,\\\label{onshellmicS} \Smic^{*}(z;\g,\t) &=
\Smic(z;\a^{*},\g,\t)\, ,\end{align}
are equal to the the corresponding gauge theory observables,
\begin{align}\label{GTR} R(z;\g,\t)  &= \Rmic^{*}(z;\g,\t)\, ,\\
\label{GTS} S(z;\g,\t)  &= \Smic^{*}(z;\g,\t)\, .\end{align}
Of course, both functions $\Rmic^{*}$ and $\Smic^{*}$ have a
complicated non-linear dependence on $\g$ that comes from solving
\eqref{micqem}.

\subsection{Outline of the paper}

The two formalisms described above have a very similar structure, with
each statement in a given framework corresponding to another statement
in the other framework. There is an obvious parallel between
\eqref{vkmac} and \eqref{ukmic}, \eqref{Zmac} and \eqref{Zmic},
\eqref{Smacdef} and \eqref{Rmicdef}, \eqref{wmacdef} and
\eqref{wmicdef}, \eqref{macqem} and \eqref{micqem}. We have indicated
in each row of the following table quantities that play analogous
r\^oles in the two formalisms. This mapping will be justified and made
more precise in the following Sections. Similar but much more detailed
tables are given at the end of the paper.

\begin{equation*}\begin{array}{|m{2.25in}|m{2.25in}|}
\hline\parbox[c]{0pt}{\rule{0pt}{6ex}}
\parbox[c]{\linewidth}{\pbs\centering{\textbf{Macroscopic formalism}}}
&\pbs\centering{\textbf{Microscopic formalism}}\\
\hline
\parbox[c]{0pt}{\rule{0pt}{4ex}}
\pbs\centering{Glueballs $\s$} & \pbs\centering{Scalars $\a$}\\
\hline
\parbox[c]{0pt}{\rule{0pt}{4ex}}
\pbs\centering{Hermitian matrix $M$} & \pbs\centering{Colored
partition $\cpart$}\\
\hline \parbox[c]{0pt}{\rule{0pt}{4ex}} \pbs\centering{Curved
background $\veps$} & \pbs\centering{$\Omega$-background $\eps$}\\
\hline \parbox[c]{0pt}{\rule{0pt}{4ex}}
\pbs\centering{Superpotential couplings $\g$} & 
\pbs\centering{Prepotential couplings $\t$}\\
\hline\parbox[c]{0pt}{\rule{0pt}{4ex}} \pbs\centering{Matrix model
partition function $\Fmac$} & \pbs\centering{Prepotential
$\Fmic$}\\
\hline\parbox[c]{0pt}{\rule{0pt}{4ex}} \pbs\centering{Macroscopic
superpotential $\wmac(\s,\g,\t)$} & \pbs\centering{Microscopic
superpotential $\wmic(\a,\g,\t)$}\\
\hline\parbox[c]{0pt}{\rule{0pt}{4ex}}
\pbs\centering{$\Smac(z;\s,\g)\, \d z$} &
\pbs\centering{$z\Rmic(z;\a,\t)\,\d z$}\\
\hline\parbox[c]{0pt}{\rule{0pt}{4ex}}
\pbs\centering{$\Rmac(z;\s,\g,\t)\,\d z$} &
\pbs\centering{$\Smic'(z;\a,\g,\t)\, \d z$}\\
\hline
\end{array}\end{equation*}

The formal structural similarities between the formalisms should not
hide the fact that the macroscopic and microscopic approaches are both
technically and conceptually very different. Clearly the matrix model
integrals at the basis of the macroscopic formalism and the sums over
colored partitions at the basis of the microscopic formalism are
totally different objects. A very important point is that the
microscopic formalism is an approach from first principles. The
equations \eqref{GTR} and \eqref{GTS} must be true by construction.
This is unlike their conjectured macroscopic analogues
\eqref{conjmacR} and \eqref{conjmacS}.

Our aim in the following will be to prove the equivalence between the
two formalisms, which can be summarized mathematically by the two
fundamental equations
\be\label{mainConj}\boxed{\begin{aligned}\Rmac^{*}(z;\g,\t) &=
\Rmic^{*}(z;\g,\t)\\ \Smac^{*}(z;\g,\t)& =
\Smic^{*}(z;\g,\t)\end{aligned}}\ee
which must be valid in all the vacua of the theory. These equations
are remarkable mathematical identities that make the link between two
seemingly unrelated starting points to perform the calculations. On
the physics side, the two completely different-looking albeit
equivalent formulations correspond to the open string (the microscopic
set-up) and the closed string (the macroscopic set-up) descriptions of
the same gauge theory.

The paper is organized as follows. In Section 2, we discuss in details
the macroscopic formalism. We establish the equivalence between the
matrix model formulas \eqref{vkmac}, \eqref{Zmac} and the generalized
Konishi anomaly equations for our extended $\nn=1$ theory
\eqref{Lag1}. We compute explicitly the functions $\Smac(z;\s,\g)$ and
$\Rmac(z;\s,\g,\t)$ from \eqref{vkmac} and \eqref{ukmac}. We then
study the critical points of $\wmac$, solving \eqref{macqem} in full
generality. The result yields explicit expressions for $\Smac^{*}$ and
$\Rmac^{*}$. In Section 3, we focus on the microscopic formalism.
Using results from Marshakov and Nekrasov \cite{MN} and the strategy
developed in \cite{mic1,mic2}, we compute explicitly $\Rmic(z;\a,\t)$
and $\Smic(z;\a,\g,\t)$. It turns out that $\Smic(z)$ is an infinitely
multi-valued analytic function, whereas the other generating functions
are always two-valued. We then solve the equations \eqref{micqem}, and
provide a full microscopic derivation of the anomaly equations which
are at the heart of the macroscopic approach. This includes the
derivation of conjectures made in \cite{mic2} about the generators of
the equations and their algebra. Equations \eqref{mainConj} then
follow. Section 4 contains our conclusions and future prospects. We
have also included two appendices. In Appendix A, we present the proof
of a generalization of the Riemann bilinear relations that plays an
important r\^ole in the main text. In Appendix B, we illustrate the
solution of the extended model \eqref{Lag1} in the particular case of
the rank one vacua.

\section{The macroscopic formalism}
\setcounter{equation}{0}
\subsection{The anomaly equations, $\boldsymbol{S}_{\mac}$ and the
matrix model}

We start from the generalized Konishi anomaly equations for the model
\eqref{Lag1}. We do not try to justify these equations beyond the
usual perturbative arguments \cite{CDSW} for the moment, since our
point of view is to develop the macroscopic formalism in this Section
using the usual hypothesis, which will be eventually proven by
comparing with the results of the microscopic approach in Section 3.

We thus follow \cite{CDSW} and consider, in perturbation theory, the
variations
\begin{align}\label{pertLn} \delta_{L_{n}} X & = -\zeta X^{n+1}\, ,\\
\label{pertJn}\delta_{J_{n}} X & = \frac{\zeta}{16\pi^{2}}
W^{\alpha}W_{\alpha} X^{n+1}\, ,\end{align}
where $\zeta$ is an infinitesimal parameter. These variations are 
generated by the operators
\be\label{LnJnpert} L_{n} = -X^{n+1}\frac{\delta}{\delta X}\,\cvp\quad
J_{n}=\frac{1}{16\pi^{2}}W^{\alpha}W_{\alpha}X^{n+1}\frac{\delta}{\delta
X}\,\cdotp\ee
They act on the observables \eqref{defu} and \eqref{defv} as
\be\label{LJact} L_{n}\cdot u_{m} = -m u_{n+m}\, ,\quad
J_{n}\cdot u_{m} = - m v_{n+m}\, ,\quad  L_{n}\cdot v_{m} =
-m v_{n+m}\, ,\quad J_{n}\cdot v_{m} = 0\ee
and satisfy the algebra
\be\label{pertalg} [L_{n},L_{m}] = (n-m) L_{n+m}\, ,\quad
[L_{n},J_{m}] = (n-m) J_{n+m}\, ,\quad [J_{n},J_{m}]=0\, .\ee
The last equation in \eqref{LJact} is a consequence of the
anticommuting nature of the chiral vector superfield $W^{\alpha}$ in
the chiral ring at the perturbative level. We refer the reader to
\cite{CDSW} for details.

Performing the changes of variables corresponding to the variations
\eqref{pertLn} and \eqref{pertJn} in the gauge theory path integral
yield the following two equations
\begin{align}\label{ano1} & -N W'(z)\Rmac(z) - N V'(z)\Smac(z) + 2
\Rmac(z)\Smac(z) + N^{2}\Delta_{R}(z) = 0\, ,\\ \label{ano2} & -N
W'(z)\Smac(z) + \Smac(z)^{2} + N^{2}\Delta_{S}(z) = 0\, .\end{align}
The polynomial terms $N^{2}\Delta_{R}$ and $N^{2}\Delta_{S}$ are
necessary to make the equations consistent with the asymptotics
\be\label{asymac}
\Smac(z)\underset{z\rightarrow\infty}{\sim}\frac{v_{0,\,\mac}}{z}\,
\cvp\quad
\Rmac(z)\underset{z\rightarrow\infty}{\sim}\frac{N}{z}\ee
that follow from the definitions \eqref{Smacdef} and \eqref{Rmacdef}.
The first two terms in the left hand side of \eqref{ano1} come from
the variation of the tree-level superpotential \eqref{treeWgen}. The
first term in \eqref{ano2} has the same origin. Note however that the
polynomial $V$ does not contribute to \eqref{ano2} because $J_{n}\cdot
v_{m} = 0$. The terms $2\Rmac\Smac$ and $\Smac^{2}$ are generated by
one-loop anomalous jacobians in the path integral, in strict parallel
with the usual one-loop Konishi anomaly \cite{konishi}.

In the perturbative framework where \eqref{ano1} and \eqref{ano2} are
derived, the variables $u_{k,\,\mac}$ and $v_{k\, ,\mac}$ must satisfy
algebraic constraints that follow from their definitions in terms of a
finite-size $N\times N$ matrix $X$. For example, there exists
polynomials $\mathscr P_{\text{pert},\, p}$ such that
\be\label{urelcl} u_{N+p,\,\mac} = \mathscr P_{\text{pert},\,
p}(u_{1,\,\mac},\ldots,u_{N,\,\mac})\, ,\quad p\geq 1\, .\ee
Similarly, only $v_{0,\,\mac},\ldots,v_{N-1,\,\mac}$ are independent.
It is not too difficult to show that \eqref{ano1} and \eqref{ano2} are
consistent with \eqref{urelcl} only if $\Rmac$ and $\Smac$ coincide
with their classical values \cite{ferchiral}. This is an unorthodox
way to rederive the standard perturbative non-renormalization theorem
for the chiral operators expectation values.

In order to carry on with the macroscopic approach, we shall use
the\smallskip\\
\newcounter{pageconj}\setcounter{pageconj}{\arabic{page}}
\textsc{Non-perturbative anomaly conjecture} \cite{ferchiral,mic2}:
\emph{The non-perturbative corrections to \eqref{ano1} and
\eqref{ano2} are such that they can be absorbed in a non-perturbative
redefinition of the variables that enter the equations.}\smallskip\\
One of the most important contribution of our work is to give in
Section 3 the first direct proof of this conjecture. For the moment we
consider it as the basic assumption of the macroscopic formalism. So
we can use \eqref{ano1} and \eqref{ano2}, but with relations
\be\label{urelq} u_{N+p,\,\mac} = \mathscr
P_{p}(u_{1,\,\mac},\ldots,u_{N,\,\mac};q,\la_{0},\ldots,\la_{d_{V}})\ee
that can be a priori arbitrary as long as they are consistent with the
symmetries \eqref{asign}.

Let us note that the non-perturbative anomaly conjecture implies that
the equations \eqref{LJact} and \eqref{pertalg} must get very strong
quantum corrections \cite{mic2}. We refer the reader to
\cite{ferchiral,mic2} and to Section \ref{anomicSec} for a more
extensive discussion of these conceptually very important points.

\begin{figure}
%\centerline{\includegraphics[width=14cm]{fig1}}
\centerline{\epsfig{file=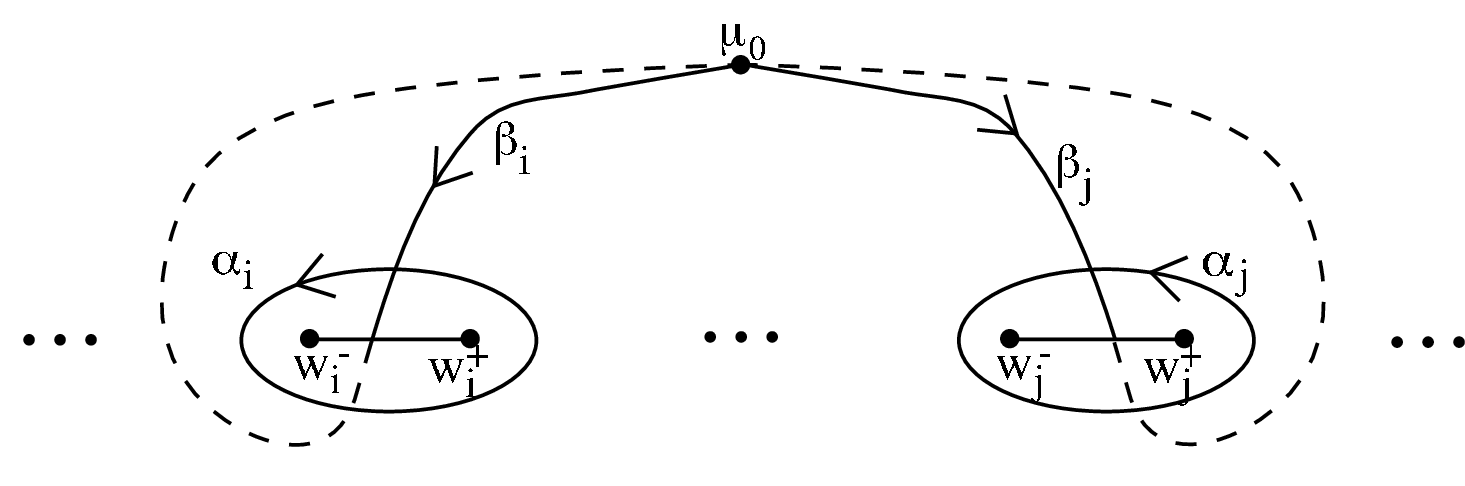,width=14cm}}
\caption{The curve $\mathcal C_{\mac,\, r}$ defined by
\eqref{Maccurve}. On the first sheet (plain lines), the asymptotic
conditions \eqref{asymac} are valid. We have depicted the contours
$\alpha_{i}$ and $\beta_{i}$ used in the main text. The contour at
infinity is given by $\alpha=\sum_{i}\alpha_{i}$. The open contours
$\beta_{i}$ start and end at $z=\mu_{0}\rightarrow\infty$, which
corresponds to the points $P_{0}$ and $Q_{0}$ on the first and second
sheets respectively.\label{fig1}}
\end{figure}

This being said, we can use \eqref{ano2} to find $\Smac$. Since the
equation does not depend on the polynomial $V$, we find that $\Smac$
does not depend on $\t$, as was claimed in Section \ref{macfor}. Thus
the function $\Smac$ is the same as in the usual $\nn=1$ gauge theory
studied in \cite{CDSW}, and in particular \eqref{vkmac} holds (it is a
direct consequence of the fact that \eqref{ano2} coincides with the
loop equation of the matrix model). Explicitly, \eqref{ano2} implies
that
\be\label{Smacf1} \Smac(z;\s,\g) = \frac{N}{2}\Bigl( W'(z) -
\sqrt{W'(z)^{2}- 4\Delta_{S}(z)}\Bigr) .\ee
The minus sign in front of the square root in \eqref{Smacf1} is found
by using the asymptotics \eqref{asymac}. The function $\Smac(z)$ is a
two sheeted function with $r\leq d_{W}$ branch cuts. The integer
$d_{W}-r$ is given by the number of double roots of the polynomial
$W'^{2}-4\Delta_{S}$,
\be\label{Macfact} W'(z)^{2} - 4\Delta_{S}(z) =
N_{d_{W}-r}(z)^{2}y_{\mac,\, r}^{2}\, ,\ee
where $N_{d_{W}-r}$ is a polynomial of degree $d_{W}-r$. We see that
$\Smac(z)$ is a meromorphic function on a genus $r-1$ hyperelliptic
curve of the form
\be\label{Maccurve}\mathcal C_{\mac,\, r}:\ y_{\mac,\, r}^{2} =
\prod_{i=1}^{r}(z-w_{i}^{-})(z-w_{i}^{+})\, .\ee
This curve, with some contours used in the main text, is depicted in
Figure \ref{fig1}. The configurations corresponding to a given value
of $r$ correspond to the description of the rank $r$ vacua of the
gauge theory. This can be straightforwardly checked by studying the
classical limit, which in this formalism corresponds to $s\rightarrow
0$. In particular, classically $w_{i}^{-}=w_{i}^{+}=w_{i}$ satisfies
$W'(w_{i})=0$.

The function $\Smac$ given by \eqref{Smacf1} depends a priori on the
$d_{W}$ undetermined parameters that enter into $\Delta_{S}$. The
factorization condition \eqref{Macfact} yields $d_{W}-r$ constraints,
and thus there remains $r$ free parameters. Using the relation between
$\Smac$ and the matrix model expectation values \eqref{vkmac}, these
$r$ free parameters can be related to the $r$ parameters $s_{i}$
defined after equation \eqref{ssirel} by
\be\label{sidef} s_{i} = \frac{1}{2i\pi
N}\oint_{\alpha_{i}}\lmac\, ,\ee
where the meromorphic one-form $\lmac$ was defined in
\eqref{macformdef}. Equations \eqref{Smacf1}, \eqref{Macfact} and
\eqref{sidef} thus give the full prescription to compute the
generating function $\Smac(z;\s,\g)$. Note that the asymptotics
\eqref{asymac} implies that
\be\label{svorel} s = \sum_{i}s_{i} = \frac{1}{2i\pi
N}\oint_{\alpha}\lmac = \frac{v_{0,\,\mac}}{N}\,\cvp\ee
or equivalently
\be\label{v0maccalc}v_{0,\,\mac}(\s,\g) =v_{0,\,\mac}(\s)=
N\sum_{i}s_{i}\, ,\ee
which is a particularly simple formula.

The fact that $\Smac$ is single-valued on the curve \eqref{Maccurve}
implies the following period integrals,
\begin{align}\label{aperSmac}\oint_{\alpha_{i}}\!
\Smac'\,\d z &= 0\, ,\\ \label{bperSmac}\int_{\beta_{i}}\!\Smac'\,\d z
&= NW'(\mu_{0})\, .\end{align}
These equations will have non-trivial counterparts in the microscopic
formalism.

\subsection{The glueball superpotential and $\boldsymbol{R}_{\mac}$}

The next important ingredient in the macroscopic formalism is the
glueball superpotential given by \eqref{wmacdef}.

\subsubsection{Consistency with the anomaly equations and
$\boldsymbol{R}_{\mac}$}
\label{corrconSec}

We are now going to check the $\g$- and $\t$-dependence of $\wmac$ by 
showing that the equations
\be\label{vkmacwmac}\begin{split}v_{0,\,\mac}(\s,\g) &=
\frac{\partial\wmac}{\partial\la_{-1}}\,\cvp\\
v_{k,\,\mac}(\s,\g) & =k\frac{\partial\wmac}{\partial\la_{k-1}}\quad
\text{for}\quad k\geq 1\end{split} \ee
and \eqref{ukmac} are consistent with the anomaly equations
\eqref{ano1} and \eqref{ano2}. This is a generalization of the
analysis made when $\la_{k}=0$ for all $k\geq 0$ in \cite{CDSW}.

The equations \eqref{vkmacwmac} are actually trivially satisfied.
Indeed, all the dependence in $\boldsymbol\la$, or equivalently in
$\t$, comes from the term containing $V$ in \eqref{wmacdef}. Using the
fact that $\lmac$ does not depend on $\t$, we see that this term is
precisely designed to be consistent with \eqref{vkmacwmac}.

Checking the consistency of \eqref{ukmac} is more interesting. Let us 
introduce the loop insertion operator
\be\label{loopop}\Lmac(z) = -\sum_{k\geq
1}\frac{k}{z^{k+1}}\frac{\partial}{\partial g_{k-1}}\,\cdotp\ee
Equations \eqref{vkFmac} and \eqref{ukmac} are equivalent to
\begin{align}\label{loopSmac} \Smac(z) & = \frac{v_{0,\,\mac}}{z} +
N\Lmac(z)\cdot\Fmac\, ,\\
\label{loopRmac} \Rmac(z) & = \frac{N}{z} - \Lmac (z)\cdot\wmac\,
.\end{align}
Applying the operator $\Lmac(z)$ on \eqref{wmacdef} and using the
above two equations together with \eqref{v0maccalc} then yields
\be\label{RmacW1} \Rmac(z) =
\frac{1}{N}\sum_{i=1}^{r}N_{i}\frac{\partial\Smac(z)}{\partial s_{i}}
- \frac{1}{2i\pi}\oint_{\alpha}\!\Lmac(z)\cdot\Smac(z')V(z')\,\d z'\,
. \ee

This formula can be written in a more elegant way. Let us introduce a
canonical basis $\{h_{i}\}_{1\leq i\leq r}$ of meromorphic one-forms
on $\mathcal C_{\mac,\, r}$, satisfying
\be\label{himacdef} \frac{1}{2i\pi}\oint_{\alpha_{j}}h_{i} =
\delta_{ij}\, ,\ee
that are holomorphic everywhere except at the points at infinity where
they may have a simple pole. Explicitly, these forms can be written as
\be\label{psiihi} h_{i} = \psi_{i}(z)\,\d z = \frac{p_{i}}{y_{\mac,\,r
}}\,\d z\, ,\ee
where the $p_{i}(z) = z^{r-1}+\cdots$ are monic polynomials of degree
$r-1$ fixed by the conditions \eqref{himacdef}. Note that these
conditions ensure that the $\{p_{i}\}_{1\leq i\leq r}$ form a basis of
the vector space of polynomials of degree at most $r-1$. By using the
explicit expression \eqref{Smacf1} for $\Smac$, we get
\be\label{derSmac1} \frac{\partial\Smac(z)}{\partial s_{i}} =
\frac{N\partial\Delta_{S}(z)/\partial s_{i}}{\sqrt{W'(z)^{2} -
4\Delta_{S}(z)}}\,\cdotp\ee
Taking the derivative of the factorization condition \eqref{Macfact},
we see that the roots of $N_{d_{W}-r}$ must also be roots of the
polynomial $\partial\Delta_{S}(z)/\partial s_{i}$, and thus
$\partial\Smac(z)/\partial s_{i}$ must be a linear combination of the 
functions $\psi_{i}$ appearing in \eqref{psiihi}. Actually, taking the
derivative of \eqref{sidef} with respect to $s_{j}$, and comparing
with \eqref{himacdef}, we find that
\be\label{derSmac}\frac{\partial\Smac(z)}{\partial s_{i}} =
N\psi_{i}(z)\, .\ee
In terms of \eqref{macformdef}, this is equivalent to
\be\label{dermacform}\frac{\partial\lmac}{\partial s_{i}} = Nh_{i}\,
.\ee
Equation \eqref{derSmac} allows to write the first term in the right
hand side of \eqref{RmacW1} in a suggestive way.

Let us now express $\Lmac(z)\cdot\Smac(z')$, by taking the derivative
of \eqref{vkmac} with respect to the couplings $\g$ using
\eqref{macmeadef}. We find
\begin{multline}\label{LmaconSmac} \Lmac(z)\cdot\Smac(z') =
\frac{N}{\veps^{2}}\biggl(
\vevsBe{\veps\!\Tr\frac{1}{z-X}\,\veps\!\Tr\frac{1}{z'-X}}\\ -
\vevsBe{\veps\!\Tr\frac{1}{z-X}}\vevsBe{\veps\!\Tr\frac{1}{z'-X}}\biggr)\,
.\end{multline}
The first term in \eqref{LmaconSmac} comes from the derivative of the
numerator in \eqref{vkmac}, and the second term comes from the
derivative of the partition function $\Zmac$ in the denumerator. We
have factored out explicitly $1/\veps^{2}$ to emphasize the fact that
in the planar $\veps\rightarrow 0$ limit, it is the combination
$\veps\!\Tr$ that has a finite limit. In particular, the right hand side
of \eqref{LmaconSmac} gets contributions from genus one, non-planar
diagrams in the matrix model Feynman graph expansion. Plugging
\eqref{LmaconSmac} and \eqref{derSmac} in \eqref{RmacW1}, we find the 
basic formula for $\Rmac$ in the macroscopic formalism,
\begin{multline}\label{RmacW} \Rmac(z;\s,\g,\t) =
\sum_{i=1}^{r}N_{i}\psi_{i}(z) - \frac{N}{\veps^{2}}\biggl(
\vevsBe{\veps\!\Tr\frac{1}{z-X}\,\veps\!\Tr V(X)}\\ -
\vevsBe{\veps\!\Tr\frac{1}{z-X}}\vevsBe{\veps\!\Tr V(X)}\biggr)\,
,\end{multline}
where the limit $\veps\rightarrow 0$ is understood. Let us note that
the constant term in $V$ does not contribute to \eqref{RmacW}, and
thus in the usual gauge theory with field-independent tree-level gauge
coupling we have the much simpler formula $\Rmac =
\sum_{i}N_{i}\psi_{i}$. Turning on non-constant terms in $V$ makes
non-planar contributions to the matrix model relevant.

Equation \eqref{RmacW} is conceptually interesting, but in order to
compare with the anomaly equation \eqref{ano1} we need a more concrete
formula. We can actually evaluate the integral in the right hand side
of \eqref{RmacW1} more explicitly. We need to use the identity
\be\label{Lmaczzp} \Lmac(z)\cdot\Smac(z') = \Lmac(z')\cdot\Smac(z)\,
,\ee
which follows either from \eqref{loopSmac} and
$[\Lmac(z),\Lmac(z')]=0$, or from the explicit expression
\eqref{LmaconSmac}. The expansions \eqref{loopop} and \eqref{Vexp}
then yield
\be\label{LmacSmacint}\begin{split}
\frac{1}{2i\pi}\oint_{\alpha}\!\Lmac(z)\cdot\Smac(z')V(z')\,\d z' &= 
\frac{1}{2i\pi}\oint_{\alpha}\!\Lmac(z')V(z')\,\d z' \cdot\Smac(z)\\
&= -\sum_{k\geq 0}\la_{k}\frac{\partial \Smac(z)}{\partial
g_{k}}\,\cvp\end{split} \ee
which implies that
\be\label{RmacW2} \Rmac(z;\s,\g,\t) = \sum_{i=1}^{r}N_{i}\psi_{i}(z) +
\sum_{k\geq 0}\la_{k}\frac{\partial \Smac(z;\s,\g)}{\partial
g_{k}}\,\cdotp\ee
The derivatives $\partial \Smac(z)/\partial g_{k}$ can be computed
from \eqref{Smacf1} in parallel with our previous computation of the
derivatives $\partial \Smac(z)/\partial s_{i}$ in \eqref{derSmac} and
above. Plugging the result in \eqref{RmacW2}, and also using the
explicit form of the $\psi_{i}$ given in \eqref{psiihi}, we find that
\be\label{Rmacfinal} \Rmac(z;\s,\g,\t) = \frac{N}{2}\biggl( V'(z) +
\frac{D_{R}(z)}{y_{\mac,\, r}}\biggr)\, ,\ee
where $D_{r}(z)$ is a polynomial of degree $d_{V}+r = \deg V'+r$. Its
coefficients are determined by the asymptotics \eqref{asymac}, which
yields $d_{V}+2$ constraints, and by the conditions
\be\label{Rmacaper} \frac{1}{2i\pi}\oint_{\alpha_{i}}\!\Rmac\,\d z 
= N_{i}\, ,\ee
which yield $r$ additional constraints, only $r-1$ of which are
independent from the previous ones because $\sum_{i}N_{i}=N$. The
period integrals \eqref{Rmacaper} follow for example from
\eqref{RmacW2} and \eqref{sidef}.

We can now easily compare this result with the prediction from the
anomaly equations. Using the solution to \eqref{ano2} in \eqref{ano1},
we find
\be\label{Rmacf1}\Rmac(z;\s,\g,\t) = \frac{N}{2}\biggl( V'(z) +
\frac{2\Delta_{R}(z) - W'(z)V'(z)}{\sqrt{W'(z)^{2} - 4
\Delta_{S}(z)}}\biggr).\ee
The anomaly equations do not put constraints on $\Delta_{R}$, and thus
$\Rmac$ in \eqref{Rmacf1} has generically $d_{W}$ branch cuts, each
corresponding to a critical point of $W$. To describe vacua which
classically correspond to having $N_{i}$ eigenvalues of $X$ sitting at
the $i^{\text{th}}$ critical point, we need to impose the constraints
\eqref{Rmacaper}, now for $1\leq i\leq d_{W}$. When only $r$ of the
$N_{i}$ are non-zero, the factorization \eqref{Macfact} must take
place, and the would-be poles of $\Rmac(z)$ at the roots of
$N_{d_{W}-r}(z)$ must vanish (otherwise the corresponding $N_{j}$
would be non-zero). The formula \eqref{Rmacf1} then reduces to
\eqref{Rmacfinal}, as was to be shown.

Let us emphasize that \eqref{Rmacaper} is non-trivial at the quantum
level \cite{ferchiral}. Indeed, there are many a priori consistent
forms for the relations \eqref{urelq} that would violate the
quantization conditions \eqref{Rmacaper}. The non-trivial statement is
that the quantization conditions must be satisfied at the quantum
level for the particular relations \eqref{urelq} for which the anomaly
equations take the simple forms \eqref{ano1} and \eqref{ano2} (see the
non-perturbative anomaly conjecture on page \arabic{pageconj}). In the
way we have presented it, these constraints \eqref{Rmacaper} look like
additional inputs that are logically independent from the anomaly
equations themselves. Remarkably, this is not the case
\cite{ferchiral, pchiral}, see the discussion at the beginning of
Section \ref{MacqemSec}.

\subsubsection{Consistency with the $\boldsymbol{\uR}$ symmetry}

A useful consistency condition on $\wmac$ comes from the $\uR$
symmetry in \eqref{asign} \cite{ferproof},
\be\label{uReq}\begin{split} \wmac(\s,\g,\t) &= \sum_{k\geq
0}g_{k}\frac{\partial\wmac}{\partial g_{k}} +
\sum_{i=1}^{r}s_{i}\frac{\partial\wmac}{\partial s_{i}}\\
& = \frac{1}{2i\pi}\oint_{\alpha}\Rmac W\,\d z +
\sum_{i=1}^{r}s_{i}\frac{\partial\wmac}{\partial s_{i}}\,
\cdotp\end{split}\ee
Comparing with \eqref{wmacdef}, we see that this a non-trivial formula
in our formalism. We are going to present a first derivation based on
matrix model identities along the lines of Section 4.2.1 of
\cite{ferproof}. Another derivation will be presented in Section
\ref{uRvisitSec}.

We start from the definition
\be\label{uRder1} N\veps\,\vevsbe{\Tr V(X)}
=\frac{1}{2i\pi}\oint_{\alpha} V\lmac =
\frac{N\veps}{\Zmac}\int\!\d\mu_{\mac}^{M}\Tr V(M)\, ,\ee
and we perform the infinitesimal variations
\be\label{sivar} \delta s_{i} = \zeta s_{i}\, ,\quad \delta\veps = \zeta 
\veps\, .\ee
Taking into account the variations of the global $\veps$ factor, of
the numerator and of the denominator in the right hand side of
\eqref{uRder1}, we obtain, in the $\veps\rightarrow 0$ limit and by
using \eqref{dermacform},
\begin{multline}\label{uRder2}
\frac{1}{2i\pi}\oint_{\alpha}\sum_{i}s_{i}V\frac{\partial\lmac}
{\partial s_{i}} =
\frac{N}{2i\pi}\oint_{\alpha}\sum_{i}s_{i}Vh_{i} =
N\veps\,\vevsbe{\Tr V(X)}\\ +
\frac{N}{\veps^{2}}\Bigl( \vevsb{\veps\!\Tr W(X)\,\veps\!\Tr V(X)} -
\vevsb{\veps\!\Tr W(X)}\vevsb{\veps\!\Tr V(X)}\Bigr).
\end{multline}
The same reasoning starting from \eqref{Zmac} shows that
\cite{ferproof}
\be\label{FmacMMid0} \sum_{i}s_{i}\frac{\partial\Fmac}{\partial s_{i}}
- 2\Fmac = \frac{1}{2i\pi N}\oint_{\alpha} W\lmac\, .\ee
Taking the derivative with respect to $s_{i}$ and using
\eqref{derSmac}, we get
\be\label{FmacMMid} -\frac{\partial\Fmac}{\partial s_{i}}
+\sum_{j}s_{j}\frac{\partial^{2}\Fmac}{\partial s_{i}\partial
s_{j}}=
\frac{1}{2i\pi}\oint_{\alpha}Wh_{i}\, .\ee
If we compute from the definition \eqref{wmacdef}, we thus find
\be\label{proofuR}\begin{split}
\sum_{i}s_{i}\frac{\partial\wmac}{\partial s_{i}}& =
\frac{N}{2i\pi}\oint_{\alpha}\sum_{i}s_{i}Vh_{i}-
\sum_{i,j}N_{j}s_{i}\frac{\partial^{2}\Fmac}{\partial s_{i}\partial
s_{j}}\\
&= \frac{1}{2i\pi}\oint_{\alpha}V\lmac -
\sum_{i}N_{i}\frac{\partial\Fmac}{\partial s_{i}} -
\frac{1}{2i\pi}\oint_{\alpha}W\sum_{i}N_{i}h_{i} \\ +&
\frac{N}{\veps^{2}}\Bigl(\vevsb{\veps\!\Tr W(X)\,\veps\!\Tr V(x)} -
\vevsb{\veps\Tr W(X)}\vevsb{\veps\Tr V(X)}\Bigr)\\
&= \wmac - \frac{1}{2i\pi}\oint_{\alpha}\Rmac W\,\d z\, .
\end{split}\ee
To go from the first to the second equality in \eqref{proofuR}, we
have used \eqref{uRder2} and \eqref{FmacMMid}, and to go from the
second to the third identity we have used \eqref{wmacdef} and
\eqref{RmacW}.

\subsection{The macroscopic quantum equations of motion}
\label{MacqemSec}

In the foregoing subsections, we have obtained explicit results for
the generating functions $\Smac(z;\s,\g)$ and $\Rmac(z;\s,\g,\t)$, see
equations \eqref{asymac}, \eqref{Smacf1}, \eqref{Macfact},
\eqref{sidef}, \eqref{Rmacfinal} and \eqref{Rmacaper}. We are now
going to compute the on-shell generating functions
$\Smac^{*}(z;\g,\t)$ and $\Rmac^{*}(z;\g,\t)$, by solving
\eqref{macqem} using the explicit formula \eqref{wmacdef}.

\subsubsection{On the consistency of the chiral ring}

Before doing that, we would like to briefly discuss the following
conceptually important question: is the formula \eqref{wmacdef} a new
axiom of the macroscopic formalism, or is it enough to postulate the
anomaly equations? We have seen in \ref{corrconSec} that the $\t$- and
$\g$-dependence of $\wmac$ is fixed by comparing with the correlators
deduced from the anomaly equations. There remains an undetermined
piece of the form $w(\s)$ in $\wmac$. Of course, this piece plays a
crucial r\^ole in solving \eqref{macqem}. Originally, it was thought
that the precise form of this term, which encodes a crucial part of
the non-perturbative gauge dynamics, needs to be postulated in
addition to the anomaly equations themselves. Remarquably, it turns
out that this is not necessary: the full $\s$-dependence of $\wmac$
follows from consistency conditions once the anomaly equations have
been postulated \cite{ferproof, ferchiral, pchiral}. The same
consistency conditions also imply that the quantization conditions
\eqref{Rmacaper} must be valid. This is a deep feature of the
macroscopic formalism. We refer the reader to \cite{pchiral} for an
extensive discussion, but let us briefly sketch the ideas involved.

For the vacua of rank $r=1$, the argument is actually very simple
\cite{ferproof}. In this case, $\wmac(s,\g,\t)$ depends on only one
variable $s$, and the full $s$-dependence is then completely fixed by
the consistency with the $\uR$ symmetry. Indeed, equation \eqref{uReq}
is not invariant if we add to $\wmac$ an arbitrary function of $s$.
The condition \eqref{Rmacaper} also follows immediately from the
asymptotic behavior \eqref{asymac}.

The case of the vacua of ranks $r>1$ is much more interesting. One of
the quantum equations of motion \eqref{macqem} is still related to the
consistency with the $\uR$ symmetry, but there remains $r-1$
independent constraints. In all the cases that have been studied (and
we are going to show that this is true in the present extended model
as well), they have the form of quantization conditions for the
compact periods of $\Rmac\d z$,
\be\label{qcRmac} \frac{1}{2i\pi}\oint_{\beta_{i}-\beta_{j}}\Rmac\,\d
z \in\mathbb Z\, .\ee
Combined with \eqref{Rmacaper}, we see that all the periods of
$\Rmac\d z$ over the non-trivial cycles of the curve $\mathcal
C_{\mac,\, r}$ are integers. Equivalently, the quantum characteristic
function
\be\label{qcf} F^{*}_{\mac}(z;\g,\t) = \bigl\langle\det
(z-X)\bigr\rangle\, ,\ee
which satisfies
\be\label{qcfrel}\frac{F^{*'}_{\mac}(z;\g,\t)}{F^{*}_{\mac}(z;\g,\t)}
= \Rmac^{*}(z;\g,\t)\, ,\ee
is well-defined (single-valued) on the curve $\mathcal C_{\mac,\, r}$,
together with $\Rmac^{*}$ and $\Smac^{*}$. The main general statement
is as follows:\smallskip\\
\textsc{Chiral ring consistency conjecture} \cite{ferchiral}:
\emph{The anomaly equations are consistent with the existence of
kinematical relations in the chiral ring of the form \eqref{urelq} if
and only if the quantization conditions \eqref{Rmacaper} and
\eqref{qcRmac} are satisfied. }\smallskip\\
Consistency is non-trivial because the relations \eqref{urelq} show
that there is only a finite number of independent variables, and the
anomaly equations yield an infinite set of constraints on this finite
set of variables.

We have proven the above conjecture in the case of the ordinary
$\nn=1$ theory with a constant $V$ and arbitrary $W$, including when
$\Nf\leq 2N$ flavors of fundamental quarks with general couplings to
$X$ are added \cite{pchiral}, and we believe that it is always true.
It shows in particular that the full $\s$-dependence of $\wmac$ is
fixed by the algebraic consistency of the chiral ring.

\subsubsection{The equations of motion}
\label{RBSec1}

We need the following standard relation for the matrix model partition
function (see for example \cite{ferCY} and references therein)
\be\label{Fmacder}\frac{\partial\Fmac}{\partial s_{i}} =
\frac{1}{N}\int_{\beta_{i}}\lmac + 2s\ln\mu_{0} - W(\mu_{0})\, .\ee
Using \eqref{wmacdef}, \eqref{dermacform} and \eqref{Rmacaper}, we
thus obtain
\be\label{derwmac1} \frac{\partial\wmac}{\partial s_{j}} =
\frac{N}{2i\pi}\oint_{\alpha}Vh_{j} -
\frac{1}{2i\pi}\sum_{i=1}^{r}\oint_{\alpha_{i}}\!\Rmac\,\d
z\int_{\beta_{i}}h_{j} - 2N\ln\mu_{0}\, .\ee
Let us introduce
\be\label{Gdwmac} H_{j}(P) = \int_{P_{0}}^{P}h_{j} + \ln\mu_{0}\, .\ee
The analytic continuation $\hat H_{j}$ of $H_{j}$ through the
$r^{\text{th}}$ branch cut is given by\footnote{See Appendix A for
details on the analytic continuation of functions like $H_{j}$.}
\be\label{GGhwmac} H_{j}(z) + \hat H_{j}(z) =
\int_{\beta_{r}}h_{j}+2\ln\mu_{0}\, ,\ee
which is found by integrating $\psi_{j}+\hat\psi_{j} = 0$ and finding
the constant of integration by looking at $z\rightarrow\infty$. Let us
now use the Riemann bilinear relation derived in Appendix \ref{Aa}
with $F=\Rmac$ and $G=H_{j}$. The formula \eqref{stspecial} applies
because $\Rmac$ is a well-defined meromorphic function on $\mathcal
C_{\mac,\, r}$. Using
\be\label{Rmachat} \Rmac(z) + \hat{R}_{\text{mac}}(z) = NV'(z)\, ,\ee
which follows from \eqref{Rmacf1}, and also \eqref{himacdef}, we find
\begin{multline}\label{RB1a}\int_{\beta_{j}}\!\Rmac\,\d z
-\frac{1}{2i\pi}
\sum_{i=1}^{r}\oint_{\alpha_{i}}\!\Rmac\,\d z\int_{\beta_{i}}h_{j} +
\frac{1}{2i\pi}\oint_{\alpha}\!\Rmac\,\d z \int_{\beta_{r}}h_{j}= \\
-\frac{1}{2i\pi}\oint_{\alpha}\biggl[\Rmac H_{j} + \Bigl(NV'-\Rmac\Bigr)
\Bigl(\int_{\beta_{r}}h_{j} + 2\ln\mu_{0}-H_{j}\Bigr)\biggr]
,\end{multline}
or equivalently
\be\label{RB1b}\int_{\beta_{j}}\!\Rmac\,\d z = \frac{1}{2i\pi}
\sum_{i=1}^{r}\oint_{\alpha_{i}}\!\Rmac\,\d z\int_{\beta_{i}}h_{j}
+2N\ln\mu_{0} + \frac{1}{2i\pi}\oint_{\alpha}\bigl(NV'H_{j} - 2\Rmac
H_{j}\bigr)\d z\, .\ee
An integration by part immediately yields
\be\label{RB1c} \frac{1}{2i\pi}\oint_{\alpha}\!V'H_{j}\,\d z
=-\frac{1}{2i\pi}\oint_{\alpha}Vh_{j} + V(\mu_{0})\, .\ee
Using the asymptotics \eqref{asymac} and
$H_{j}(z)\smash{\underset{z\rightarrow\infty}{\sim}} \ln z$, we also get
\be\label{RB1d} \frac{1}{2i\pi}\oint_{\alpha}\!\Rmac H_{j}\,\d z =
N\ln\mu_{0}\, .\ee
Putting \eqref{derwmac1}, \eqref{RB1b}, \eqref{RB1c} and \eqref{RB1d} 
together, we finally obtain
\be\label{derwmacfi} \frac{\partial\wmac}{\partial s_{i}} =
-\int_{\beta_{i}}\!\Rmac\,\d z + NV(\mu_{0})-2 N\ln\mu_{0}\, . \ee

The above formula is natural when one uses the parameters $\t$.
However, from the physics point of view, a small refinement is needed,
because $t_{1}$ is not really a good parameter. The good parameter is
the instanton factor \eqref{qdef} $q=e^{t_{1}}$, and the theory must
be invariant when $t_{1}\rightarrow t_{1}+2i\pi$, which corresponds to
a $2\pi$ shift of the $\vartheta$ angle. From \eqref{wmacdef}, we know
that $t_{1}$ enters $\wmac$ only through a term $s t_{1} = s\ln q$.
Thus we find that vacua related to each other by $2\pi$ shifts of the
$\vartheta$ angle are described by different macroscopic
superpotentials of the form $\wmac+2i\pi k s$, for $k\in\mathbb Z$.
The most general vacua are thus obtained by looking at solutions of
the equations
\be\label{derwmacfinal} \int_{\beta_{i}}\!\Rmac^{*}\,\d z =
NV(\mu_{0})-2 N\ln\mu_{0}+2i\pi\mathbb Z\, .\ee
As a last remark, let us note that \eqref{qcRmac} follows as a special
case of \eqref{derwmacfinal}.

\subsubsection{The $\boldsymbol{\uR}$ symmetry revisited}
\label{uRvisitSec}

It is instructive to rederive \eqref{uReq} starting from
\eqref{derwmacfi}. Using \eqref{wmacdef}, \eqref{sidef} and
\eqref{Fmacder}, \eqref{uReq} is equivalent to
\begin{multline}\label{uReqbis}
\sum_{i=1}^{r}\biggl[\oint_{\alpha_{i}}\!\Rmac\,\d
z\int_{\beta_{i}}\!\Smac\,\d z - \oint_{\alpha_{i}}\!\Smac\,\d
z\int_{\beta_{i}}\!\Rmac\,\d z\biggr] =\\ N\oint_{\alpha}\bigl(\Smac V -
\Rmac W\bigr)\d z + 2i\pi N^{2}\bigl(W(\mu_{0}) - s V(\mu_{0})\bigr)
.\end{multline}
We let the reader check that this is a consequence of
\eqref{stspecial}, with $F=\Smac$ and $G=\rho$ with
\be\label{rhomacdef} \rho(P) = \int_{P_{0}}^{P}\!\Rmac\,\d z
+N\ln\mu_{0}\, .\ee
Useful formulas to perform the check are
\be\label{Smachat} \Smac(z) + \hat{S}_{\mac}(z) = NW'(z)\, ,\ee
which follows from \eqref{Smacf1}, and 
\be\label{rrhatmac}\rho(z) + \hat\rho(z) = NV(z) +
\int_{\beta_{r}}\!\Rmac\,\d z - NV(\mu_{0}) + 2N\ln\mu_{0}\, , \ee
which follows by integrating \eqref{Rmachat}. One also needs
\be\label{rhoche} \frac{1}{2i\pi}\oint_{\alpha}\!\Smac\rho\,\d z 
= N^{2}s\ln\mu_{0}\, ,\ee
which is similar to \eqref{RB1d}.

\subsection{The solution in the macroscopic formalism}

Let us summarize our findings. The functions $\Smac^{*}$ and
$\Rmac^{*}$ are fully determined by the constraints
\be\boxed{\begin{gathered}\label{macsol}
-N W'(z)\Rmac^{*}(z) - N V'(z)\Smac^{*}(z) + 2
\Rmac^{*}(z)\Smac^{*}(z) + N^{2}\Delta_{R}(z) = 0\, ,\\
-N W'(z)\Smac^{*}(z) + \Smac^{*}(z)^{2} + N^{2}\Delta_{S}(z) = 0\\
\Smac^{*}(z)\underset{z\rightarrow\infty}{\sim}\frac{Ns}{z}\,
\cvp\quad
\Rmac^{*}(z)\underset{z\rightarrow\infty}{\sim}\frac{N}{z}\,\cvp\\
 \oint_{\alpha_{i}}\!\Rmac^{*}\,\d z \in 2i\pi\mathbb Z\,
,\quad \int_{\beta_{i}}\!\Rmac^{*}\,\d z - NV(\mu_{0}) + 2N\ln\mu_{0}\in
2i\pi\mathbb Z \, .\end{gathered}}\ee
The first three lines in \eqref{macsol} follow from the corresponding
formulas \eqref{ano1}, \eqref{ano2} and \eqref{asymac} for $\Rmac$ and
$\Smac$, while the constraints in the fourth line follow from
\eqref{Rmacaper} and \eqref{derwmacfinal}. The last equation is valid
only on-shell, i.e.\ when $\Rmac=\Rmac^{*}$ and $\Smac=\Smac^{*}$.
Note that the asymptotics (third line) and the quantization conditions
(fourth line) in \eqref{macsol} provide enough constraints to fix the
solution uniquely, up to a discrete ambiguity corresponding to the
existence of a discrete set of vacua.

The quantization conditions in \eqref{macsol} are equivalent to the
fact that the quantum characteristic function \eqref{qcf},
\eqref{qcfrel} is single-valued on the same hyperellictic curve as
$\Smac^{*}$ and $\Rmac^{*}$, with the asymptotics
\be\label{QCFasy} F^{*}_{\mac}(z) \underset{z\rightarrow\infty}{\sim}
z^{N}\, ,\quad \hat{F}^{*}_{\mac}(z)\underset{z\rightarrow\infty}{\sim}
\frac{e^{NV(z)}}{z^{N}}\,\cdotp\ee
The function $F$ has in general an essential singularity at $Q_{0}$.
It is only in the case of the non-extended theory, for which $V$ is a
constant, that it is a meromorphic function together with $\Smac^{*}$ and
$\Rmac^{*}$.

We refer the reader to Appendix \ref{Ab}, where the explicit solution
for rank one vacua is discussed.

\section{The microscopic formalism}
\setcounter{equation}{0}

In this Section, we are going to solve the model using the microscopic
formalism \cite{mic1}.

\subsection{Marshakov-Nekrasov and $\boldsymbol{\Rmic}$}

We start by computing the generating function $\Rmic$ which, in the
microscopic formalism, is the simplest object. The main property that 
we are going to use is that $\Rmic$ does \emph{not} depend on $\g$,
\be\label{rmicg}\Rmic(z;\a,\g,\t) = \Rmic(z;\a,\t)\, .\ee
This property, which is the analogue of the fact that $\Smac$ does not
depend on $\t$ in the macroscopic set-up, is manifest on the
definition \eqref{ukmic} and follows immediately from the localization
formula \cite{fucito,mic2}. We can thus compute $\Rmic$ in the $\g=0$
theory, which has $\nn=2$ supersymmetry. Fortunately for us, the
extended $\nn=2$ theory, for arbitrary $\t$, has been studied recently
by Marshakov and Nekrasov in \cite{MN}, and so we can just borrow the 
result from them.\footnote{There is an unfortunate typo in Marshakov
and Nekrasov that pollutes many of their formula. The typo first
appears in their equation (3.2), where $\t'$ should be replaced by
$\frac{1}{2}\t'$. We have corrected all their subsequent equations
accordingly.} 

The function $\Rmic(z)$ turns out to be a meromorphic function on the 
genus $N-1$ hyperelliptic curve
\be\label{SWcurve} \mathcal C_{\mic}:\ y^{2} =
\prod_{i=1}^{N}(x-x_{i}^{-})(x-x_{i}^{+})\, .\ee
On $\mathcal C_{\mic}$, we define marked points $P_{0}$ and $Q_{0}$,
as well as contours $\alpha_{i}$, $\alpha=\sum_{i}\alpha_{i}$,
$\beta_{i}$ and $\delta_{i}=\beta_{i}-\beta_{N}$ is a way similar to
what we have done on $\mathcal C_{\mac,\, r}$ in Figure
\ref{fig1}.\footnote{We use the same names for the contours on
different curves. Which curve we are referring to is always clear from
the context.} We have
\be\label{Rmic}\Rmic(z;\a,\t) = \frac{N}{2}\Bigl(V'(z) +
\frac{E_{R}(z)}{y}\Bigr)\,\cvp\ee
where $E_{R}$ is a polynomial of degree $N+\deg V'=N+d_{V}$. The
$N+d_{V}+1$ coefficients in $E_{R}$ and the $2N$ branching points of
the curve \eqref{SWcurve} are determined by the following conditions
\cite{MN}:
\begin{align}\label{asymic1}&\Rmic(z)
\underset{z\rightarrow\infty}{\sim}\frac{N}{z}\,\cvp\\
\label{micper1}
&\frac{1}{2i\pi}\oint_{\alpha_{i}}\!\Rmic\,\d z = 1\, ,\\ 
&\label{micper2} \int_{\beta_{i}}\!\Rmic\,\d z = NV(\mu_{0}) -
2N\ln\mu_{0}+2i\pi\mathbb Z\, ,\\
\label{aidef} & a_{i} = \frac{1}{2i\pi}\oint_{\alpha_{i}}\lmic\,
.\end{align}
To write these equations, we have assumed that the curve
\eqref{SWcurve} is not degenerate. This is always the case for large
enough $|a_{i}-a_{j}|$, and the solution is then uniquely specified
for any values of the $a_{i}$ by analytic continuation in the
$\a$-space. The one-form $\lmic$ is defined in \eqref{micformdef}. The
first constraint \eqref{asymic1} yields $d_{V}+2$ conditions, the
second constraint \eqref{micper1} yields $N-1$ conditions that are
independent from the previous ones, and the third and fourth
constraints \eqref{micper2} and \eqref{aidef} yield $N+N$ new
conditions, for a total of $3N+d_{V}+1$ independent conditions as
needed. Let us note that the contours $\beta_{i}$ are really defined
modulo an integral linear combination of the $\alpha_{j}$, which
explains the term $+2i\pi\mathbb Z$ in \eqref{micper2}.

The difference between the macroscopic and microscopic formalisms is
here manifest. Equation \eqref{micper2} is valid \emph{off-shell},
i.e.\ for any values of the parameters $\a$. The analogous equation
\eqref{derwmacfinal} for $\Rmac$ is valid only \emph{on-shell}, i.e.\
for the particular values $\s^{*}$ of the parameters $\s$ that make
$\wmac$ extremal. This is a very general feature: ``simple'' equations
in a given formalism, valid off-shell, correspond to ``complicated''
equations in the other formalism, valid only on-shell. This is the
mechanism that will allow to identify $\Rmac^{*}$ and $\Rmic^{*}$ on
the one hand and $\Smac^{*}$ and $\Smic^{*}$ on the other hand, in
spite of the fact that the off-shell functions have very different
properties.

\subsection{The generating function $\boldsymbol{\Smic}$}
\label{SmicSec}

Let us now compute the generating function for the generalized
glueball operators $\Smic$. This function has been studied in details 
in \cite{mic2}. In particular, a general formula for the
$v_{k,\,\mic}$ was derived (the derivation was made in the usual
theory with $t_{k}=0$ for $k\geq 2$, but it actually applies without
change to the case of the extended theory),
\begin{multline}\label{vnform} v_{k,\,\mic}(\a,\g,\t,\eps) =
\frac{N}{(k+1)(k+2)}\frac{1}{\eps^{2}}\Bigl( \vevab{\Tr W(X) \Tr
X^{k+2}}_{\eps}\\ - \vevab{\Tr W(X)}_{\eps}\vevab{\Tr
X^{k+2}}_{\eps}\Bigr)\, .\end{multline}
This can be conveniently rewritten in terms of $\Smic$ as
\begin{multline}\label{SmicV}\Smic''(z;\a,\g,\t) = \frac{N}{\eps^{2}}
\biggl(
\vevaBe{\Tr\frac{1}{z-X}\,\Tr W(X)}\\ -
\vevaBe{\Tr\frac{1}{z-X}}\vevaBe{\Tr W(X)}\biggr)\,
,\end{multline}
where the ${}'$ always means the derivative with respect to $z$. This
formula, which is the microscopic analogue of \eqref{RmacW}, will be
rederived shortly.\footnote{Note that \eqref{RmacW} and \eqref{SmicV}
are actually valid at finite $\veps$ and $\eps$ respectively, even
though we are only interested in the $\veps\rightarrow 0$ and
$\eps\rightarrow 0$ limits in the present paper.} It shows that the
subleading terms in the small $\eps$ expansion contribute to the
glueball operators. Another contribution of \cite{mic2} was to compute
explicitly $\Smic$ up to two instantons using \eqref{SmicV} and the
definitions given in \ref{micfor}. We are going to derive in the
present Section the \emph{exact} formula for $\Smic$, by imitating the
macroscopic derivation of $\Rmac$ in \ref{corrconSec}.

Let us start by explaining the origin of \eqref{SmicV} in the present
set-up. We shall need a few simple identities. We note that
\eqref{ukpart} implies that $u_{1,\,\cpart}$ does not depend on the
colored partition $\cpart$,
\be\label{u1part} u_{1,\,\cpart} = \sum_{i=1}^{N}a_{i}\, .\ee
From this we deduce that
\be\label{u1simple} u_{1,\,\mic}(\a,\t) = u_{1,\,\mic}(\a) =
\sum_{i=1}^{N}a_{i} = a\, ,\ee
which is a particularly simple formula, as well as
\be\label{spfacu1} \vevabe{\Tr X\Tr X^{k}} = \vevabe{\Tr X}\vevabe{\Tr
X^{k}}\ee
for any $k\geq 0$. We introduce the microscopic ``loop insertion
operator''
\be\label{loopmicop}\Lmic(z)
=-\frac{1}{z}\frac{\partial}{\partial\la_{-1}} -\sum_{k\geq
1}\frac{k}{z^{k+1}}\frac{\partial}{\partial\la_{k-1}}\,\cvp \ee
or equivalently
\be\label{loopmicop2}\Lmic''(z) = -N \sum_{k\geq
2}\frac{k}{z^{k+1}}\frac{\partial}{\partial t_{k-1}}\,\cdotp \ee
Note that the derivatives are taken at $\a$ fixed here, whereas in the
similar macroscopic operator \eqref{loopop} the derivatives are at
$\s$ fixed. This point should be clear and we shall not repeat it in
the following. Using \eqref{u1simple}, equations \eqref{ukFmic} and
\eqref{vkmic} are then equivalent to
\begin{align}\label{loopRmic} \Rmic(z) & = \frac{N}{z}
+\frac{a}{z^{2}}  - \frac{2}{N}\,
\Lmic''(z)\cdot\Fmic\, ,\\
\label{loopSmic} \Smic(z) & = - \Lmic (z)\cdot\wmic\, .\end{align}
Applying the operator $\Lmic(z)$ on \eqref{wmicdef} and using
\eqref{loopSmic} then yields
\be\label{SmicW1} \Smic(z) = -
\frac{1}{2i\pi}\oint_{\alpha}\!\Lmic(z)\cdot\Rmic(z')W(z')\,\d z'\, .
\ee
On the other hand, if we take the derivative of \eqref{ukmic} with
respect to the couplings $\t$ or $\boldsymbol{\la}$ using
\eqref{munurel} and \eqref{micmeadef} and taking into account
\eqref{spfacu1}, we get
\begin{multline}\label{LmiconRmic} \Lmic''(z)\cdot\Rmic(z') =
-\frac{N}{\eps^{2}}\biggl(
\vevaBe{\Tr\frac{1}{z-X}\Tr\frac{1}{z'-X}}\\ -
\vevaBe{\Tr\frac{1}{z-X}}\vevaBe{\Tr\frac{1}{z'-X}}\biggr)\,
.\end{multline}
Plugging \eqref{LmiconRmic} into \eqref{SmicW1}, we obtain
\eqref{SmicV}.

To get a more explicit result, we use
\be\label{Lmiczzp} \Lmic''(z)\cdot\Rmic(z') =
\Lmic''(z')\cdot\Rmic(z)\, , \ee
which yields
\be\label{LmicRmicint}\begin{split}
\frac{1}{2i\pi}\oint_{\alpha}\!\Lmic''(z)\cdot\Rmic(z')W(z')\,\d z' &= 
\frac{1}{2i\pi}\oint_{\alpha}\!\Lmic''(z')W(z')\,\d z' \cdot\Rmic(z)\\
&= -N\sum_{k\geq 1}g_{k}\frac{\partial \Rmic(z)}{\partial
t_{k}}\,\cdotp\end{split} \ee
Equation \eqref{SmicW1} thus implies that 
\be\label{SmicW2} \Smic''(z;\a,\g,\t) = N\sum_{k\geq
1}g_{k}\frac{\partial \Rmic(z;\a,\t)}{\partial t_{k}}\,\cdotp\ee
This important result shows that $\Smic''(z)$ is a meromorphic
function on the curve $\mathcal C_{\mic}$ \eqref{SWcurve}, since
$\Rmic(z)$ and thus also its variations have this property. We can
actually go one step further. Using \eqref{SmicW2}, the derivatives of
equations \eqref{micper1} and \eqref{micper2} with respect to the
$t_{k}$s indeed imply that
\be\label{Sppper} \oint_{\alpha_{i}}\!\Smic''\,\d z = 0\, ,\quad
\oint_{\delta_{i}}\!\Smic''\,\d z = 0\, .\ee
This shows that
\be\label{Smicpdef}\Smic'(P) = \int_{P_{0}}^{P}\!\Smic''\,\d z\ee
is also a well-defined meromorphic function on $\mathcal C_{\mic}$,
since \eqref{Sppper} ensures that $\Smic'(P)$ does not depend on the
path chosen to perform the integral in \eqref{Smicpdef}. 

Another way to see this, and to gain more information on the pole
structure of $\Smic'$, is to use the microscopic quantum
characteristic function
\be\label{Fmic}F_{\mic}(P) = \vevab{\det(z-X)} =
\mu_{0}^{N}\exp\int_{P_{0}}^{P}\!\Rmic\,\d z\, .\ee
It is a single-valued function on $\mathcal C_{\mic}$ thank's to
\eqref{micper1} and \eqref{micper2}. It is actually holomorphic
everywhere except at the points at infinity. It has a pole of order
$N$ at $P_{0}$ and an essential singularity at $Q_{0}$,
\be\label{QCFmicasy} F_{\mic}(z) \underset{z\rightarrow\infty}{\sim}
z^{N}\, ,\quad \hat F_{\mic}(z)\underset{z\rightarrow\infty}{\sim}
\frac{e^{NV(z)}}{z^{N}}\,\cdotp\ee
We have used as usual the hatted notation to indicate the value of a
function on the second sheet of the hyperelliptic curve on which it is
defined. Let us remark that \eqref{QCFmicasy} and \eqref{QCFasy}
coincide, not surprisingly, but \eqref{QCFmicasy} is valid for any
$\a$ whereas \eqref{QCFasy} is valid only at $\s=\s^{*}$. The most
general form for $F_{\mic}$ that is compatible with the above
constraints is
\be\label{Fmicgef}F_{\mic}(z) = \phi_{1}(z) + \phi_{2}(z) y\, ,\ee
for some entire functions $\phi_{1}$ and $\phi_{2}$ on the complex
plane whose asymptotics can be found from \eqref{QCFmicasy}. Moreover,
it is manifest from the definition \eqref{Fmic} and the explicit
formula \eqref{Rmic} for $\Rmic$ that $F_{\mic}$ can never vanish at
finite $z$. Thus we also have
\be\label{Fmicgef2}\frac{1}{F_{\mic}(z)} = \varphi_{1}(z) +
\varphi_{2}(z) y\, ,\ee
for some entire functions $\varphi_{1}$ and $\varphi_{2}$. Let us now
consider objects of the form
\be\label{fdelta} f_{\delta}(z) = \frac{\delta
F_{\mic}}{F_{\mic}} = \delta\ln F_{\mic}\, ,\ee
where $\delta$ represents the derivative with respect to a parameter,
for example the $t_{k}$ or the $a_{i}$. Clearly $f_{\delta}$ is a
meromorphic function on $\mathcal C_{\mic}$, because $\delta\ln\hat
F_{\mic}\smash{\underset{z\rightarrow\infty}{\sim}}N\delta V(z)$ has
at most an ordinary pole at $Q_{0}$. Moreover, it follows from
\eqref{SWcurve} that
\be\label{deltay} \delta y = \frac{\rho_{\delta}}{y}\ee
for some polynomial $\rho_{\delta}$. Putting \eqref{Fmicgef},
\eqref{Fmicgef2}, \eqref{fdelta} and \eqref{deltay} together, we find
that
\be\label{fdform} f_{\delta}(z) = p_{\delta}(z) +
\frac{q_{\delta}(z)}{y}\,\cdotp\ee
The entire functions $p_{\delta}$ and $q_{\delta}$ must actually be
polynomials because $f_{\delta}$ is meromorphic.

If we apply this remark to the primitive of \eqref{SmicW2}, which
reads
\be\label{SpmicW}\Smic'(z;\a,\g,\t) = N\sum_{k\geq
1}g_{k}\frac{\partial\ln F_{\mic}(z;\a,\t)}{\partial t_{k}}\,\cvp\ee
we deduce that $\Smic'$ must take the form
\be\label{Smicpa} \Smic'(z) =\frac{N}{2}\Bigl( p(z) +
\frac{E_{S}(z)}{y}\Bigr) \ee
for some polynomials $p$ and $E_{S}$ (the overall factor has been chosen
for convenience). If we now use
\be\label{RRhatmic} \Rmic(z) + \hat{R}_{\mic}(z) = NV'(z)\, ,\ee
that follows from \eqref{Rmic}, and \eqref{SmicW2}, we find
\be\label{SSpphatmic} \Smic''(z) + \hat{S}_{\mic}''(z) = NW'''(z)\, .\ee
Integrating, we obtain
\be\label{SSphatmic} \Smic'(z) + \hat{S}_{\mic}'(z) = NW''(z) + c =
Np(z)\, .\ee
The constant of integration $c$ is found to vanish by looking at the
limit $z=\mu_{0}\rightarrow\infty$ and using
\be\label{betarSpp}\int_{\beta_{i}}\!\Smic''\,\d z = NW''(\mu_{0})\,
,\ee
which is a straightforward consequence of \eqref{SmicW2} and
\eqref{micper2}. Overall, we have thus obtained
\be\label{Smicp} \Smic'(z) =\frac{N}{2}\Bigl( W''(z) +
\frac{E_{S}(z)}{y}\Bigr). \ee
The polynomial $E_{S}$ is determined in the following way. First, from
the asymptotics
\be\label{asymicS} \Smic'(z)
\underset{z\rightarrow\infty}{\sim}-\frac{v_{0,\,\mic}}{z^{2}}\,\cvp\ee
we deduce that $\deg E_{S} = N+\deg W'' = N+d_{W}-1$. The condition
\eqref{asymicS} actually puts $d_{W}+1$ constraints on the
coefficients of $E_{S}$, which leaves $N-1$ unknown. The missing
constraints, to be derived shortly, are given by
\be\label{micSpera}\oint_{\alpha_{i}}\!\Smic'\,\d z = 0\, .\ee
This is the microscopic analogue to \eqref{Rmacaper}. Equations
\eqref{Smicp}, \eqref{asymicS} and \eqref{micSpera} give a full
prescription to compute exactly $\Smic'$, and thus all the generalized
glueball operators in the microscopic formalism. In particular, we
have checked the result successfully with the explicit calculations
made in \cite{mic2}.

To finish, let us prove \eqref{micSpera}. The idea is to integrate by
part and then to use \eqref{SmicW2} and \eqref{micformdef},
\be\label{comaperS} \oint_{\alpha_{i}}\!\Smic'\,\d z =
-\oint_{\alpha_{i}}\! z\Smic''\,\d z = -N\sum_{k\geq
1}g_{k}\frac{\partial}{\partial t_{k}}\oint_{\alpha_{i}}\lmic\, .\ee
The result then follows by taking the derivative of \eqref{aidef} with
respect to $t_{k}$, which yields zero. It is natural to ask about the
$\delta_{i}$-periods of $\Smic'$. Clearly, they will \emph{not} vanish
in general, and thus $\Smic$ is \emph{not} well-defined on the curve
\eqref{SWcurve}. Actually, it is an infinitely-many valued function of
$z$. This is very different from the macroscopic function $\Smac$,
which is always two-valued because it satisfies the anomaly equation
\eqref{ano2}. For the equality $\Smac^{*}=\Smic^{*}$ to be valid, the
microscopic equations of motion \eqref{micqem} must put constraints on
the $\beta_{i}$-periods of $\Smic'\d z$. This is what we are going to
study now.

\subsection{The microscopic quantum equations of motion}

We define, in strict parallel with what was done in Section
\ref{corrconSec} for the curve \eqref{Maccurve}, a canonical basis of
one-forms $\{h_{i}\}_{1\leq i\leq N}$ on $\mathcal C_{\mic}$,
\be\label{hipermic}\frac{1}{2i\pi}\oint_{\alpha_{j}}h_{i} =
\delta_{ij}\, ,\ee
holomorphic everywhere except at the points at infinity where they may
have simple poles. Explicitly,
\be\label{psiimic} h_{i} = \psi_{i}\,\d z = \frac{p_{i}}{y}\,\d z\,
,\ee
where the $p_{i}(z) = z^{N-1}+\cdots$ are monic polynomials of degree 
$N-1$ fixed by the conditions \eqref{hipermic}. We shall need the
identity
\be\label{derlmic}\frac{\partial\lmic}{\partial a_{i}} = h_{i} -
\d (z\psi_{i})\, .\ee
To derive this equation, let us write
\be\label{lmicnew}\lmic = z\Rmic\,\d z = -\ln F_{\mic}\,\d z + \d
(z\ln F_{\mic})\, .\ee
From \eqref{fdelta} and \eqref{fdform} applied to $\delta =
\partial/\partial a_{i}$, we know that
\be\label{daFmic}\frac{\partial\ln F_{\mic}}{\partial a_{i}} = p(z) + 
\frac{q(z)}{y}\ee
for some polynomials $p$ and $q$. Integrating \eqref{RRhatmic} with
respect to $z$ and then taking the derivative with respect to $a_{i}$,
we find that necessarily $p=0$. Moreover, 
\be\label{asymFoF} \frac{1}{F_{\mic}}\frac{\partial F_{\mic}}{\partial
a_{i}} = \mathcal O(1/z)\, ,\ee
and thus $\deg q = N-1$. This shows that $\partial\ln
F_{\mic}/\partial a_{i}$ is a linear combination of the $\psi_{i}$.
Comparing \eqref{hipermic} with the derivative of \eqref{aidef} with
respect to $a_{j}$, we actually find
\be\label{daFmicbis}\frac{\partial\ln F_{\mic}}{\partial a_{i}} =
-\psi_{i}\ee
which, using \eqref{lmicnew}, is equivalent to \eqref{derlmic}.

The definition \eqref{wmicdef} yields
\be\label{derwmica}\frac{\partial\wmic}{\partial a_{i}} =
\frac{1}{2i\pi}\oint_{\alpha}\frac{W}{z}\frac{\partial\lmic}{\partial 
a_{i}}\,\cdotp\ee
Using \eqref{derlmic} and an integration by part, we thus obtain
\be\label{derwmicb} \frac{\partial\wmic}{\partial a_{i}}
=\frac{1}{2i\pi}\oint_{\alpha}W'h_{i}\, .\ee

We now apply the Riemann bilinear relation \eqref{stspecial} with
$F=\Smic'$ and $g=h_{i}$. The calculation is strictly similar to the
one done in Section \ref{RBSec1}, and thus we shall be brief. Using
\eqref{micSpera}, we get, using the same definition as in
\eqref{Gdwmac},
\be\label{RBmic1}\begin{split} \int_{\beta_{i}}\!\Smic'\,\d z& =
-\frac{1}{2i\pi}\oint_{\alpha}
\biggl[\Smic' H_{i} + \Bigl(NW''-\Smic'\Bigr)
\Bigl(\int_{\beta_{r}}h_{i} + 2\ln\mu_{0}-H_{i}\Bigr)\biggr]\\
& = NW'(\mu_{0}) - \frac{N}{2i\pi}\oint_{\alpha}W'h_{i}\, .
\end{split}\ee
Equation \eqref{derwmicb} is thus equivalent to
\be\label{derwmic}\frac{\partial\wmic}{\partial a_{i}} =
-\frac{1}{N}\int_{\beta_{i}}\!\Smic'\,\d z + W'(\mu_{0})\, .\ee
This is exactly what the discussion at the end of \ref{SmicSec} was
suggesting.  On-shell, we have
\be\label{onshellSmic}\int_{\beta_{i}}\!\Smic^{*'}\,\d z =
NW'(\mu_{0})\, .\ee
In particular, $\oint_{\delta_{i}}\Smic^{*'}\d z = 0$ which, together
with \eqref{micSpera}, implies that $\Smic^{*}$ is a meromorphic
function on \eqref{SWcurve}.

It is important to note that the curve $\eqref{SWcurve}$ can
degenerate (and actually must always do so if $\deg W'<N$) on the
solutions to \eqref{derwmic}. A very explicit discussion of how this
happens is given in \cite{mic1}. To understand better this point, let
us look in more details into the consequences of \eqref{onshellSmic}.
Since $\Smic^{*}$ is two-sheeted, we can integrate \eqref{SSphatmic}
(remember that \eqref{betarSpp} implies that $c=0$) to get
\be\label{Sonshat} \Smic^{*}(z) + \hat{S}_{\mic}^{*}(z) = N W'(z) +
\tilde c\, .\ee
Looking at the limit $z=\mu_{0}\rightarrow\infty$, we obtain
\be\label{ctildecomp} \int_{\beta_{i}}\!\Smic^{*'}\,\d z =
NW'(\mu_{0}) + \tilde c\ee
and thus the equation of motion \eqref{onshellSmic} implies that
$\tilde c = 0$. The function $\Smic^{*}(NW' - \Smic^{*})$ is thus
single-valued and meromorphic, meaning that it must be a rational
function of $z$. Poles for finite values of $z$ at points where $z$ is
a good coordinate on the curve (i.e.\ excluding the branching points)
would yield similar poles for $\Smic^{*'}$ which we know are not
present, see \eqref{Smicp}. Similarly, a pole of order $n$ at a
branching point yields after taking the derivative with respect to $z$
a pole of order $n+2$ for $\Smic^{*'}$ (because $\d y/\d z\propto
1/y$). This is impossible because \eqref{Smicp} shows that
$\Smic^{*'}$ only has simple poles at the branching points. Finally,
this discussion implies that
\be\label{polsolS} \Smic^{*}(z)\bigl(NW'(z) - \Smic^{*}(z)\bigr) =
N^{2}\Delta^{*}_{S,\,\mic}(z)\ee
is a polynomial whose degree is fixed to be $d_{W}-1$ by the
asymptotics $\Smic^{*}=\mathcal O(1/z)$ at infinity. Solving this
quadratic equation we find that
\be\label{Smicf1} \Smic^{*}(z;\g,\t) = \frac{N}{2}\Bigl( W'(z) -
\sqrt{W'(z)^{2}- 4\smash[t]{\Delta^{*}_{S,\,\mic}}(z)}\Bigr) .\ee
Consistency with the fact that $\Smic^{*}$ is also defined on
\eqref{SWcurve} implies that the following factorization conditions
must be satisfied,
\begin{align}\label{micfa1} &y^{2} = M_{N-r}(z)^{2}y_{\mic,\, r}^{2}\,
,\\ \label{micfa2} & W'(z)^{2} - 4\Delta_{S,\,\mic}^{*}(z) =
N_{d_{W}-r}(z)^{2}y_{\mic,\, r}^{2}\, ,\end{align}
for some polynomials $M_{N-r}$ and $N_{d_{W}-r}$ of degrees $N-r$ and
$d_{W}-r$ respectively. A priori, the integer $r$ can take any value
consistent with the above equations. We see that both $\Smic^{*}$ and
$\Rmic^{*}$ are defined on a genus $r-1$ reduced curve
\be\label{Miccurve} \mathcal C_{\mic,\, r}:\ y_{\mic,\, r}^{2} =
\prod_{i=1}^{r}(z-v_{i}^{-})(z-v_{i}^{+})\ee
which is obtained from \eqref{SWcurve} by joining some of the branch
cuts. For example, if the $i^{\text{th}}$ and $j^{\text{th}}$ branch
cuts join, then the resulting cut has a $\beta$-type contour
$\beta_{i}'=\beta_{i}=\beta_{j}$ and a $\alpha$-type contour
$\alpha_{i}'=\alpha_{i}+\alpha_{j}$. Reshuffling the indices
appropriately and renaming the contours $\alpha_{i}$ and $\beta_{i}$
on the reduced curve to match the notations used in the macroscopic
formalism, we see that \eqref{micper1} and \eqref{micper2} become
\begin{align}\label{onpermic1}& \oint_{\alpha_{i}}\!\Rmic^{*}\,\d z
\in 2i\pi\mathbb Z\, ,\\
\label{onpermic2}&\int_{\beta_{i}}\!\Rmic^{*}\,\d z - NV(\mu_{0}) +
2N\ln\mu_{0}\in 2i\pi\mathbb Z\, ,\end{align}
on the reduced curve. The integer appearing on the right hand side of 
\eqref{onpermic1} is simply given by the number of cuts of the
original curve $\mathcal C_{\mic}$ that have joined to form the
$i^{\text{th}}$ cut of the reduced curve $\mathcal C_{\mic,\, r}$.

The link with the macroscopic formalism is almost complete. To finish
the proof, we need to show that $\Rmic^{*}$ and $\Smic^{*}$ are
related to each other consistently with the anomaly equation
\eqref{ano1} that relates $\Rmac^{*}$ and $\Smac^{*}$ (note that
\eqref{polsolS} already shows that $\Smic^{*}$ satisfies an equation
like \eqref{ano2}). We could do that by carefully analyzing the
properties of $\Rmic^{*}$ and $\Smic^{*}$, but the most elegant route
is to perform a full microscopic analysis of the anomaly equations.
This is the subject of the next Section.

\subsection{The microscopic derivation of the anomaly equations}
\label{anomicSec}

At the perturbative level, the anomaly equations are generated by the
operators \eqref{LnJnpert}. Their action on the chiral operators is
given by \eqref{LJact} and their algebra by \eqref{pertalg}. As
discussed at length in \cite{mic2}, these operators must undergo very
strong non-perturbative quantum corrections. In particular, at the
non-perturbative level, $L_{n}\cdot u_{m}$, $J_{n}\cdot u_{m}$,
$L_{n}\cdot v_{m}$ and $J_{n}\cdot v_{m}$ are not single-valued
functions of the $u_{p}$s and $v_{q}$s, and moreover the algebra
generated by $L_{n}$ and $J_{m}$ does not close.

In \cite{mic2}, it was conjectured that the non-perturbative operators
are given by\footnote{In \cite{mic2}, only the usual theory with $V$
constant was considered, but the formulas for $L_{n}$ and $J_{n}$
straightforwardly generalize to the case of arbitrary $\t$.}
\begin{align}\label{conjLn} L_{n}& =
-\frac{1}{2i\pi}\sum_{i=1}^{N}\oint_{\alpha_{i}}z^{n+1}
\Rmic(z;\a,\t)\,\d z\,\frac{\partial}{\partial
a_{i}}\,\cvp\\\label{conjJn} J_{n}& =
-\frac{1}{2i\pi}\sum_{i=1}^{N}\oint_{\alpha_{i}}z^{n+1}
\Smic(z;\a,\g,\t)\,\d z\,\frac{\partial}{\partial
a_{i}}\,\cdotp\end{align}
The fundamental requirement is that they generate the correct anomaly 
polynomials
\begin{align}\label{actionL}N L_{n}\cdot\wmic (\a,\g,\t)& = \mathscr
A_{n}(\a,\g,\t)\, ,\\ \label{actionJ}N J_{n}\cdot\wmic (\a,\g,\t)& =
\mathscr B_{n}(\a,\g,\t)\, ,\end{align}
with
\begin{align}\label{anomic1}\mathscr A_{n} & = - N\sum_{k\geq 0}
g_{k}u_{n+k+1,\,\mic}- N\sum_{k\geq 0} \la_{k}v_{n+k+1,\,\mic} +
2\sum_{k_{1}+k_{2} = n} u_{k_{1},\,\mic}v_{k_{2},\,\mic}\, , \\
\label{anomic2} \mathscr B_{n} & = - N\sum_{k\geq 0}
g_{k}v_{n+k+1,\,\mic} + \sum_{k_{1}+k_{2} = n}
v_{k_{1},\,\mic}v_{k_{2},\,\mic}\, .
\end{align}
Equivalently, in terms of the generating function we have 
\begin{align}\label{Anomic1}\begin{split}\mathscr
A(z;\a,\g,\t)&=\sum_{n\geq -1}\frac{\mathscr A_{n}(\a,\g,\t)}{z^{n+2}}
= NL(z)\cdot\wmic(\a,\g,\t)
\\& = -NW'(z;\g)\Rmic(z;\a,\t)-NV'(z;\t)\Smic(z;\a,\g,\t)\\& \hskip
1.5cm+ 2 \Rmic(z;\a,\t)\Smic(z;\a,\g,\t) +
N^{2}\Delta_{R,\,\mic}(z;\a,\g,\t)\, ,\end{split}\\
\label{Anomic2}\begin{split}\mathscr B(z;\a,\g,\t)&=\sum_{n\geq
-1}\frac{\mathscr
B_{n}(\a,\g,\t)}{z^{n+2}}=NJ(z)\cdot\wmic(\a,\g,\t)\\& =
-NW'(z;\g)\Smic(z;\a,\g,\t) \\&\hskip 3cm+
\Smic(z;\a,\g,\t)^{2}+N^{2}\Delta_{S,\,\mic}(z;\a,\g,\t)\,
,\end{split}
\end{align}
for some polynomials $\Delta_{R,\,\mic}(z)$ and
$\Delta_{S,\,\mic}(z)$. Equations \eqref{actionL} and \eqref{actionJ}
were checked in \cite{mic2} up to two instantons. Using the technology
developed in the previous Sections, we can now prove these equations
independently of the small $q$ expansion.

\subsubsection{The action of $\boldsymbol{L_{n}}$}

We use \eqref{conjLn} and \eqref{derwmic} to write
\be\label{mica1} NL_{n}\cdot\wmic =
\frac{1}{2i\pi}\sum_{i}\oint_{\alpha_{i}}\! z^{n+1}\Rmic\,\d
z\int_{\beta_{i}}\!\Smic'\,\d z - NW'(\mu_{0})u_{n+1,\,\mic}\, .\ee
The right-hand side of this equation can be evaluated by using the
Riemann bilinear relation \eqref{stspecial}, with $p=z^{n+1}$,
$F=\Rmic$ and $G=\Smic$. Using \eqref{RRhatmic},
\be\label{SShatmic} \Smic(z) + \hat{S}_{\mic}(z) = NW'(z) +
\int_{\beta_{r}}\!\Smic'\,\d z - NW'(\mu_{0})\, ,\ee
which is obtained from \eqref{SSphatmic} in a way similar to
\eqref{rrhatmac}, and \eqref{micSpera} we get
\be\label{mica2}\begin{split} NL_{n}\cdot\wmic &=
-NW'(\mu_{0})u_{n+1,\,\mic} +
u_{n+1,\,\mic}\int_{\beta_{r}}\!\Smic'\,\d z\\ &\hskip -2cm +
\frac{1}{2i\pi}\oint_{\alpha}\! z^{n+1}\biggl[\Rmic\Smic +
\Bigl(NV'-\Rmic\Bigr)\Bigl(NW'-\Smic-N W'(\mu_{0}) +
\int_{\beta_{r}}\!\Smic'\,\d z\Bigr)\biggr]\\ &=
\frac{1}{2i\pi}\oint_{\alpha}z^{n+1}\bigl(-NW'\Rmic - NV'\Smic +
2\Rmic\Smic\bigr)\d z =\mathscr A_{n}\, . \end{split}\ee
We have thus derived the first anomaly equation \eqref{actionL}.

\subsubsection{The action of $\boldsymbol{J_{n}}$}

The definition \eqref{conjJn} yields
\be\label{mica3} NJ_{n}\cdot\wmic =
\frac{1}{2i\pi}\sum_{i}\oint_{\alpha_{i}}\! z^{n+1}\Smic\,\d
z\int_{\beta_{i}}\!\Smic'\,\d z - NW'(\mu_{0})v_{n+1,\,\mic}\, .\ee
To perform the calculation, we apply the Riemann bilinear relation
with $F=G=\Smic$. In this case, the full power of the generalized
relation \eqref{stfinal} derived in the Appendix is needed, because
$F=\Smic$ is multi-valued on the curve \eqref{SWcurve} as explained at
the end of Section \ref{SmicSec}. Because $F=G$, the net effect of the
additional terms in \eqref{stfinal} with respect to the more
conventional formula \eqref{stspecial} is a crucial global factor of
2,
\be\label{mica4}\begin{split}&
2\times \frac{1}{2i\pi}\sum_{i} \oint_{\alpha_{i}}\!
z^{n+1}\Smic\,\d z \int_{\beta_{i}}\!\Smic'\,\d z = 2\times
v_{n+1,\,\mic}\int_{\beta_{r}}\!\Smic'\,\d z\\& \hskip 2cm+
\frac{1}{2i\pi}\oint_{\alpha} \! z^{n+1}\biggl[\Smic^{2} +
\Bigl(NW'-\Smic-N W'(\mu_{0}) + \int_{\beta_{r}}\!\Smic'\,\d
z\Bigr)^{2}\biggr]\\ &\hskip 3cm
= \frac{1}{i\pi}\oint_{\alpha}z^{n+1}\bigl(-NW'\Smic +
\Smic^{2}\bigr)\d z + 2N W'(\mu_{0})v_{n+1,\,\mic}\, .
\end{split}\ee
Plugging this result in \eqref{mica3}, we obtain \eqref{actionJ} as we
wished, completing the full microscopic derivation of the generalized
Konishi anomaly equations.
\subsection{The solution in the microscopic formalism}

When the quantum equations of motion \eqref{micqem} are satisfied, we 
have automatically
\be\label{onshellmicano} L_{n}\cdot\wmic = 0\, ,\quad J_{n}\cdot\wmic 
= 0\, ,\ee
and thus $\Rmic^{*}$ and $\Smic^{*}$ satisfy the anomaly equations. We
can thus summarize our findings as follows. The functions $\Rmic^{*}$
and $\Smic^{*}$ are fully determined by the constraints
\be\boxed{\begin{gathered}\label{micsol}
-N W'(z)\Rmic^{*}(z) - N V'(z)\Smic^{*}(z) + 2
\Rmic^{*}(z)\Smic^{*}(z) + N^{2}\Delta_{R,\,\mic}(z) = 0\, ,\\
-N W'(z)\Smic^{*}(z) + \Smic^{*}(z)^{2} + N^{2}\Delta_{S,\,\mic}(z) = 0\\
\Rmic^{*}(z)\underset{z\rightarrow\infty}{\sim}\frac{N}{z}\, \cvp\quad
\Smic^{*}(z)\underset{z\rightarrow\infty}{\sim}\frac{v_{0,\,\mic}}{z}
\,\cvp\\
 \oint_{\alpha_{i}}\!\Rmic^{*}\,\d z \in 2i\pi\mathbb Z\,
,\quad \int_{\beta_{i}}\!\Rmic^{*}\,\d z - NV(\mu_{0}) + 2N\ln\mu_{0}\in
2i\pi\mathbb Z \, .\end{gathered}}\ee
The first two constraints are the anomaly equations that we have just
derived. They are valid only on-shell, which is unlike their
macroscopic counterparts \eqref{ano1} and \eqref{ano2} which are valid
off-shell. The constraints in the third line of \eqref{micsol} are the
usual asymptotics that follow from the definitions of $\Rmic$ and
$\Smic$. The last constraints in the fourth line of \eqref{micsol}
correspond to \eqref{onpermic1} and \eqref{onpermic2}. The solution is
then fixed up to the usual discrete ambiguity corresponding to the
existence of a discrete set of vacua.

The \emph{on-shell} solutions of the macroscopic formalism
\eqref{macsol} and of the microscopic formalism \eqref{micsol} are
clearly identical. The fundamental result \eqref{mainConj} is thus
proven.

\section{Conclusions}
\setcounter{equation}{0}

Generalizing the work of Nekrasov and collaborators
\cite{nekrasova,MN,nekrasovb} from $\nn=2$ to $\nn=1$, we have
provided a microscopic derivation, from first principles, of the exact
results in supersymmetric gauge theories, following the ideas
explained in \cite{mic1}. The main highlight is to show that the
microscopic approach based on the Nekrasov's sums over colored
partitions and the macroscopic approach based on the Dijkgraaf-Vafa
matrix model are equivalent. A particularly interesting application is
the non-perturbative derivation of the generalized Konishi anomaly
equations.

The macroscopic and microscopic formalisms have striking structural
similarities. This is illustrated in the following tables, where each
equation in one formalism is associated to a similar equation in the
other formalism.

\begin{equation*}\begin{array}{|m{2.75in}|m{2.75in}|}
\hline
\parbox[c]{0pt}{\rule{0pt}{5ex}}
\parbox[c]{\linewidth}{\pbs\centering{\textbf{Macroscopic formalism}}}
&\pbs\centering{\textbf{Microscopic formalism}}\\
\hline
\parbox[c]{0pt}{\rule{0pt}{4ex}}
\pbs\centering{$v_{k,\,\mac}(\s,\g)=N\veps\vevsb{\Tr
X^{k}}$} & \pbs\centering{$
u_{k,\,\mic}(\a,\t)=\vevab{\Tr X^{k}}$}\\
\hline
\parbox[c]{0pt}{\rule{0pt}{4ex}}
\pbs\centering{$\lmac=\Smac(z;\s,\g)\, \d z$} &
\pbs\centering{$\lmic=z\Rmic(z;\a,\t)\,\d z$}\\
\hline
\parbox[c]{0pt}{\rule{0pt}{6ex}}
\pbs\centering{$\displaystyle\s=\frac{1}{2i\pi N}\oint_{\alpha}\lmac$}
&
\pbs\centering{$\displaystyle\a=\frac{1}{2i\pi}\oint_{\alpha}\lmic$}\\
\hline
\parbox[c]{0pt}{\rule{0pt}{6ex}}
\pbs\centering{$\displaystyle\frac{\partial\lmac}{\partial
s_{i}}=Nh_{i}$} &
\pbs\centering{$\displaystyle\frac{\partial\lmic}{\partial
a_{i}}=h_{i}-\d(z\psi_{i})$}\\
\hline
\parbox[c]{0pt}{\rule{0pt}{12ex}}
\pbs\centering{$\begin{aligned}&\wmac(\s,\g,\t) =
\frac{1}{2i\pi}\oint_{\alpha}\Smac V\,\d z \\&\qquad\qquad\qquad\qquad
- \sum\nolimits_{i}N_{i}\frac{\partial\Fmac}{\partial
s_{i}}\end{aligned}$} & \pbs\centering{$\wmic(\a,\g,\t) =\displaystyle
\frac{1}{2i\pi}\oint_{\alpha}\Rmic W\,\d z$}\\
\hline
\parbox[c]{0pt}{\rule{0pt}{12.75ex}}
\pbs\centering{$\begin{aligned}&\sum\nolimits_{i}s_{i}
\frac{\partial\wmac}{\partial s_{i}} =\\&\hskip 1cm
-\frac{1}{2i\pi}\oint_{\alpha}\Rmac W\,\d z
+\wmac \end{aligned}$} &
\pbs\centering{$\begin{aligned}&\sum\nolimits_{i}a_{i}
\frac{\partial\wmac}{\partial a_{i}} =-2v_{0,\,\mic}\\&
+\frac{1}{2i\pi}\oint_{\alpha}\bigl(zW'\Rmic + zV'\Smic\bigr)\d z
\end{aligned}$}\\
\hline
\parbox[c]{0pt}{\rule{0pt}{7ex}}
\pbs\centering{$\displaystyle u_{k,\,\mac}(\s,\g,\t) =
k\frac{\partial\wmac}{\partial g_{k-1}}$} &
\pbs\centering{$\displaystyle v_{k,\,\mic}(\a,\g,\t) =
\frac{N}{k+1}\frac{\partial\wmic}{\partial t_{k+1}}$}\\
\hline
\parbox[c]{0pt}{\rule{0pt}{7ex}}
\pbs\centering{$\displaystyle v_{k,\,\mac}(\s,\g) =\frac{N}{k+1}
\frac{\partial\wmac}{\partial t_{k+1}}$} &
\pbs\centering{$\displaystyle u_{k,\,\mic}(\a,\t) =
k\frac{\partial\wmic}{\partial g_{k-1}}$}\\
\hline
\parbox[c]{0pt}{\rule{0pt}{7ex}}
\pbs\centering{$\displaystyle v_{k,\,\mac}(\s,\g) =-Nk
\frac{\partial\Fmac}{\partial g_{k-1}}\,\cvp\ k\geq 1$} &
\pbs\centering{$\displaystyle u_{k,\,\mic}(\a,\t) =2k
\frac{\partial\Fmic}{\partial t_{k-1}}\,\cvp\ k\geq 2$}\\
\hline
\parbox[c]{0pt}{\rule{0pt}{17ex}}
\pbs\centering{$\begin{aligned}& u_{k,\,\mac}(\s,\g,\t)
=\frac{1}{2i\pi}\oint_{\alpha}\! z^{k}
\sum\nolimits_{i}N_{i}h_{i} \\& - \frac{N}{\veps^{2}}\Bigl[
\vevsbe{\veps\!\Tr X^{k}\,\veps\!\Tr V(X)}\\& -
\vevsbe{\veps\!\Tr X^{k}}\vevsbe{\veps\!\Tr V(X)}\Bigr]\end{aligned}$} &
\pbs\centering{$\begin{aligned}&v_{k,\,\mic}(\a,\g,\t)=
\frac{N}{(k+1)(k+2)}\times\\&
\frac{1}{\eps^{2}}\Bigl[\vevabe{\Tr X^{k+2}\Tr W(X)} \\& - \vevabe{\Tr
X^{k+2}}\vevabe{\Tr W(X)}\Bigr]\end{aligned}$}\\
\hline
\parbox[c]{0pt}{\rule{0pt}{11ex}} \pbs\centering{$\begin{aligned}
&\Rmac(z;\s,\g,\t) = \sum\nolimits_{i}N_{i}\psi_{i}(z)\\&\hskip 1.5cm+
\sum\nolimits_{k\geq 0}\la_{k}\frac{\partial \Smac(z;\s,\g)}{\partial
g_{k}}\end{aligned}$} &
\pbs\centering{$\begin{aligned}&\Smic''(z;\a,\g,\t) =\\ &\hskip 1.5cm N
\sum\nolimits_{k\geq
1}g_{k}\frac{\partial\Rmic(z;\a,\t)}{\partial t_{k}}\end{aligned}$}\\
\hline
\parbox[c]{0pt}{\rule{0pt}{7ex}} \pbs\centering{
$\displaystyle\frac{1}{2i\pi}\oint_{\alpha_{i}}\!\Smac'\,\d z=0$} &
\pbs\centering{$\displaystyle\frac{1}{2i\pi}\oint_{\alpha_{i}}\!\Rmic\,\d
z\in\mathbb Z$}\\
\hline
\parbox[c]{0pt}{\rule{0pt}{7ex}} \pbs\centering{
$\displaystyle\int_{\beta_{i}}\!\Smac'\,\d z= NW'(\mu_{0})$} &
\pbs\centering{$\displaystyle\int_{\beta_{i}}\!\Rmic\,\d
z=NV(\mu_{0})-2N\ln\mu_{0}$}\\
\hline
\end{array}\end{equation*}
\begin{equation*}\begin{array}{|m{2.75in}|m{2.75in}|}
\hline
\parbox[c]{0pt}{\rule{0pt}{5ex}}
\parbox[c]{\linewidth}{\pbs\centering{\textbf{Macroscopic formalism}}}
&\pbs\centering{\textbf{Microscopic formalism}}\\
\hline
\parbox[c]{0pt}{\rule{0pt}{7ex}} \pbs\centering{
$\displaystyle\frac{1}{2i\pi}\oint_{\alpha_{i}}\!\Rmac\,\d z\in\mathbb
Z$} & \pbs\centering{$\displaystyle\frac{1}{2i\pi}\oint_{\alpha_{i}}\!
\Smic'\,\d z=0$}\\
\hline
\parbox[c]{0pt}{\rule{0pt}{10ex}} \pbs\centering{$\begin{aligned}
\frac{\partial\wmac}{\partial s_{i}}& =-\int_{\beta_{i}}\!\Rmac\,\d z\\
&\hskip 1.25cm- 2N\ln\mu_{0}+NV(\mu_{0})\end{aligned}$} &
\pbs\centering{$\displaystyle \frac{\partial\wmic}{\partial a_{i}}
=-\frac{1}{N}\int_{\beta_{i}}\!\Smic'\,\d z + W'(\mu_{0})$}
\\
\hline
\parbox[c]{0pt}{\rule{0pt}{7ex}}
\pbs\centering{$\begin{aligned}-NW'\Rmac - NV'\Smac + 2\Rmac\Smac\\
+ N^{2}\Delta_{R} = 0 \end{aligned}$} &
\pbs\centering{$\begin{aligned}&NL(z)\cdot\wmic = -NW'\Rmic\\& -
NV'\Smic + 2\Rmic\Smic + N^{2}\Delta_{R,\,\mic}\end{aligned}$}\\
\hline
\parbox[c]{0pt}{\rule{0pt}{6.75ex}}
\pbs\centering{$\begin{aligned}-NW'\Smac + \Smac^{2} + N^{2}\Delta_{S}
= 0 \end{aligned}$}& \pbs\centering{$\begin{aligned}&NJ(z)\cdot\wmic =
-NW'\Smic\\&\hskip 3cm + \Smic^{2}+ N^{2}\Delta_{S,\,\mic}
\end{aligned}$}\\
\hline
\parbox[c]{0pt}{\rule{0pt}{3.75ex}}
\pbs\centering{$\boldsymbol{\Rmac^{*}=R\, ,\quad \Smac^{*}=S}$}&
\pbs\centering{$\boldsymbol{\Rmic^{*}=R\, ,\quad \Smic^{*}=S}$}\\
\hline
\end{array}\end{equation*}

A basic property of the correspondence between the formalisms is that
an off-shell, or ``kinematical'' relation on one side is typically
only valid on-shell, or ``dynamically'' on the other side. When both
sides are put on-shell, they yield equivalent results. This is
reminiscent of the electric/magnetic duality, which exchanges Bianchi
identities with equations of motion, and seems to be a common feature
of many non-trivial dualities. Actually, the duality we have been
discussing is directly related to the open/closed string duality, the
open string description corresponding to the microscopic formalism and
the closed string description to the macroscopic formalism.

There are many ways to generalize the present work. Each particular
$\nn=1$ gauge theory, with given gauge group and matter content, can
be studied along the lines of our work, and a nice equivalence between
the associated macroscopic and microscopic formalisms should follow. 

A particularly interesting avenue of research is to consider
deformations of the ordinary gauge theories, by turning on various
backgrounds. For example, we can study the theory in a non-zero
$\Omega$-background. Many ``microscopic'' formulas straightforwardly
generalize to this case. In particular, the microscopic superpotential
is still given by \eqref{wmicdef}, because the parameter $\eps$ is not
charged under the $\uR$ symmetry. \emph{The duality discussed in the
present work must generalize to the deformed theory}. In particular,
there should exist a deformation of the macroscopic formalism
corresponding to turning on $\eps$, and the on-shell equivalence
between the formalisms should hold to any order in
$\eps$.\footnote{Actually, the most general $\Omega$-background is
characterized by two complex parameters $\eps_{1}$ and $\eps_{2}$, and
we have studied the special case $\eps=\eps_{1}=\eps_{2}$. The duality
should extend to the most general case as well.} Another interesting
deformation is the gravitational background discussed in \cite{DS},
which corresponds to the non-zero parameter $\veps$ on the macroscopic
side. This deformation has not been studied yet on the microscopic
side. Understanding fully these extended dualities is likely to
involve interesting physics and mathematics. They are highly
non-trivial, yet probably fully solvable, examples of open/closed
string dualities.

\subsection*{Acknowledgements}

I would like to thank Vincent Wens and Stanislav Kuperstein for useful
discussions.

This work is supported in part by the belgian Fonds de la Recherche
Fondamentale Collective (grant 2.4655.07), the belgian Institut
Interuniversitaire des Sciences Nucl\'eaires (grant 4.4505.86), the
Interuniversity Attraction Poles Programme (Belgian Science Policy)
and by the European Commission FP6 programme MRTN-CT-2004-005104 (in
association with V.\ U.\ Brussels). Frank Ferrari is on leave of
absence from Centre National de la Recherche Scientifique, Laboratoire
de Physique Th\'eorique de l'\'Ecole Normale Sup\'erieure, Paris,
France.

% Appendix
\renewcommand{\thesection}{\Alph{section}}
\renewcommand{\thesubsection}{\arabic{subsection}}
\renewcommand{\theequation}{A.\arabic{equation}}
\setcounter{section}{0}
\section{Generalized Riemann bilinear relations}
\label{Aa}

In this Appendix, we derive an interesting generalization of the
Riemann bilinear relations that we have used again and again in the
main text.

\begin{figure}
%\centerline{\includegraphics[width=5in]{fig2}}
\centerline{\epsfig{file=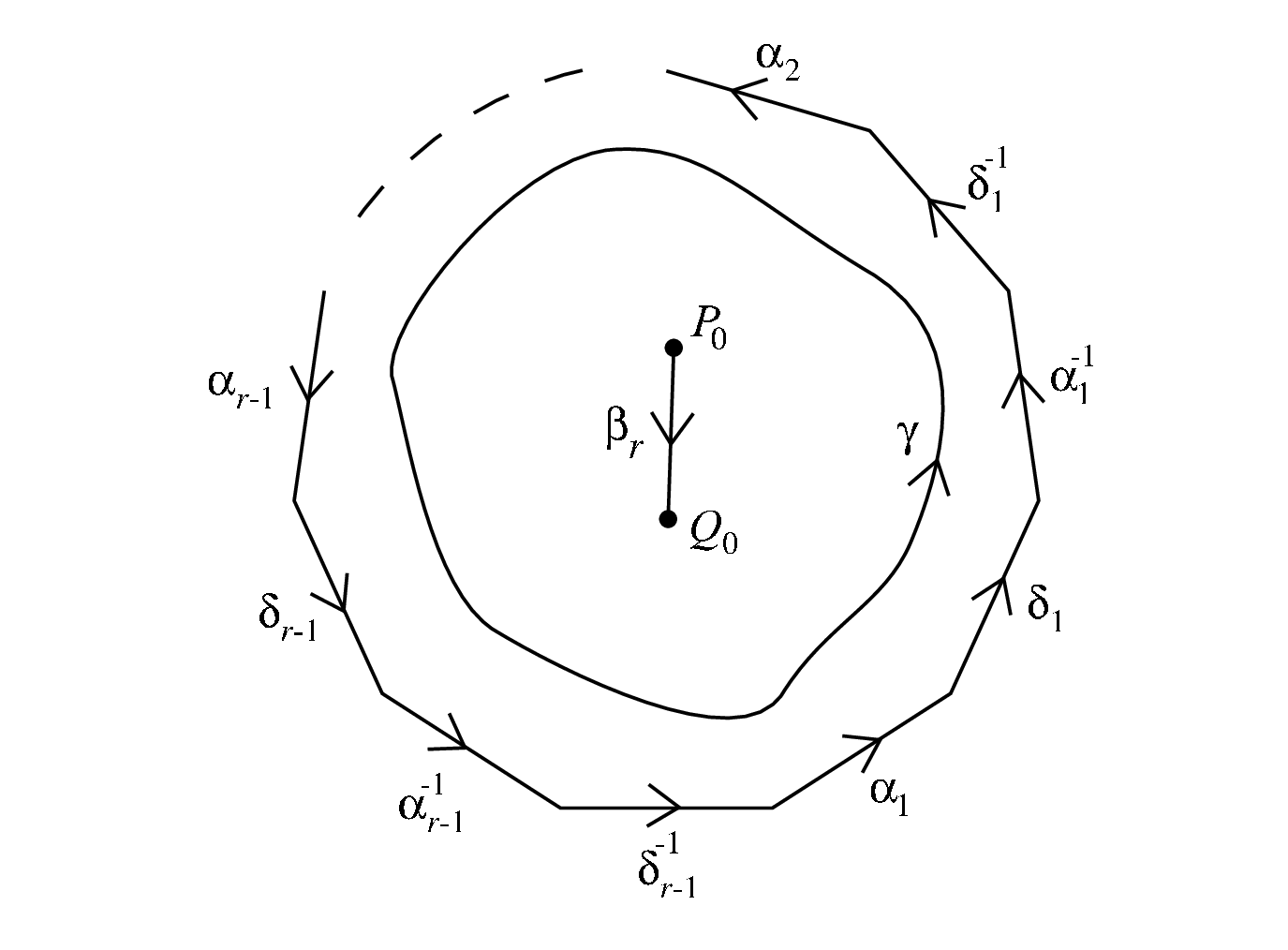,width=4in}}
\caption{The representation of the surface \eqref{Apcurve} as a
$4(r-1)$-gon with suitable identifications on the boundary. We have
depicted the contours $\alpha_{i}$, $\delta_{i}=\beta_{i}-\beta_{r}$,
$\beta_{r}$ and $\gamma$, as well as the two marked points at infinity
$P_{0}$ and $Q_{0}$.}\label{figA}
\end{figure}

We consider a hyperelliptic curve
\be\label{Apcurve} \mathcal C:\ y^{2} =
\prod_{i=1}^{r}(z-z_{i}^{-})(z-z_{i}^{+})\ee
with two marked points $P_{0}$ and $Q_{0}$ corresponding to
$z=\infty$. This curve, which can be conveniently represented as a
polygon with suitable identifications on the boundary, is depicted in
Figure \ref{figA}. The contours $\alpha_{i}$ can be chosen to encircle
the branch cuts $[z_{i}^{-},z_{i}^{+}]$, and the contours $\beta_{i}$
join $P_{0}$ to $Q_{0}$ by going through $[z_{i}^{-},z_{i}^{+}]$. The
contours $\delta_{i}$ are then defined by
$\delta_{i}=\beta_{i}-\beta_{r}$ for $1\leq i\leq r-1$, and we note
$\alpha=\sum_{i}\alpha_{i}$ the contour at infinity. It is convenient
to introduce a regulator $\mu_{0}$. In formulas containing $\mu_{0}$,
it is always understood that the points $P_{0}$ and $Q_{0}$ correspond
to $(y=\mu_{0}^{r},z=\mu_{0})$ and $(y=-\mu_{0}^{r},z=\mu_{0})$
respectively and that the limit $\mu_{0}\rightarrow\infty$ must be
taken. If $h$ is a meromorphic function on $\mathcal C$, we denote by
$h(z)$ its value on the first sheet (which we choose to be the sheet
containing $P_{0}$) and $\hat h(z)$ its value on the second sheet.

We consider two meromorphic functions $f$ and $g$ on $\mathcal C$ that
are holomorphic everywhere except at the points at infinity where they
may have poles of arbitrary order (our discussion can be
straightforwardly generalized when poles at finite $z$ are present,
but this is not needed for the applications in the main text). Let us
first assume that
\be\label{perf} \oint_{\alpha_{i}}\! f\,\d z = 0\, ,\quad
\oint_{\delta_{i}}\!f\,\d z = 0\, .\ee
Let us choose a base point $O$ on $\mathcal C$, distinct from $P_{0}$
or $Q_{0}$. A primitive $F$ of $f$ is defined by
\be\label{defF} F(P) = \int_{O}^{P}\!f\, \d z\, .\ee
The conditions \eqref{perf} ensure that $F(P)$ does not depend on the
path from $O$ to $P$ chosen to perform the integral in \eqref{defF}.
This means that $F$ is single-valued on the curve $\mathcal C$. In
particular, we can talk about the values $F(z)$ and $\hat F(z)$. It is
actually a meromorphic function on $\mathcal
C\setminus\{P_{0},Q_{0}\}$. If $f\d z$ has a non-zero residue $f_{0}$
at $P_{0}$ (and thus also a non-zero residue $-f_{0}$ at $Q_{0}$),
then $F$ has a logarithmic branch cut running from $P_{0}$ to $Q_{0}$
across which it jumps by $2i\pi f_{0}$. We shall always choose this
branch cut to go along the contour $\beta_{r}$.

Let us now waive the hypothesis \eqref{perf}. The function $F$ is then
no longer single-valued on $\mathcal C$ and we need to specify the
contour from $O$ to $P$ in \eqref{defF}. We shall always choose this
contour to lie entirely in the interior of the polygon of Figure
\ref{figA}, never going to the boundary. If $z$ is fixed, $F$ can then
take two values $F(z)$ and $\hat F(z)$ in the polygon, modulo the
$2i\pi f_{0}$ ambiguity due to the possible logarithmic cut. In the
description involving the two sheets of the curve \eqref{Apcurve},
$F(z)$ correspond to the value obtained by doing the integral
\eqref{defF} following a contour that never circles around a branch
cut $[z_{i}^{-},z_{i}^{+}]$, whereas the value $\hat F(z)$ is obtained
by doing the analytic continuation following straight a path that goes
through the same branch cut as $\beta_{r}$.

Let us also define the primitive $G$ of $g$ by
\be\label{defG} G(P) = \int_{O}^{P}\!g\, \d z\, ,\ee
following exactly the same procedure as for $F$. We then consider the 
integral
\be\label{Idef}\mathscr I = \oint_{\gamma}pFG\, \d z\, ,\ee
where $p(z)$ is an arbitrary polynomial and the contour $\gamma$ is
defined in Figure \ref{figA}. We could of course absorb the polynomial
$p$ by redefining $F$ or $G$, but it is convenient to present the
results in this form for our purposes. Let us emphasize that we do
\emph{not} assume that relations like \eqref{perf} hold for $f$ or for
$g$. We are going to compute this integral in two different ways, and
this will yield the generalization of the Riemann bilinear relations
that we are seeking.

Let us first deform the contour $\gamma$ so that it merges with the
boundary of the polygon. If the function $FG$ were single valued on
the curve $\mathcal C$, we would automatically find zero due to the
cancellation between the terms coresponding to the integral over
$\alpha_{i}$ and then $\alpha_{i}^{-1}$, and over $\delta_{i}$ and
then $\delta_{i}^{-1}$. However, to go from the contour $\alpha_{i}$
to the contour $\alpha_{i}^{-1}$, we have to follow $\delta_{i}$,
which induces a discontinuity $\oint_{\delta_{i}}\!f\d z$ for $F$ and
$\oint_{\delta_{i}}\!g\d z$ for $G$. The same phenomenon occurs when we
go from the contour $\delta_{i}$ to $\delta_{i}^{-1}$: following
$\alpha_{i}^{-1}$, we pick discontinuities $-\oint_{\alpha_{i}}\!f\d
z$ and $-\oint_{\alpha_{i}}\!g\d z$ for $F$ and $G$ respectively.
Overall we thus obtain
\be\label{st1a} \begin{split}\mathscr I &=\sum_{i=1}^{r-1}\biggl[
\oint_{\alpha_{i}}\biggl(pFG - p\Bigl(F +\oint_{\delta_{i}}\! f\d
u\Bigr)\Bigl(G+\oint_{\delta_{i}}\!g\d u\Bigr)\biggr)\d z\\ &\hskip 
3cm +
\oint_{\delta_{i}}\biggl(pFG - p\Bigl(F -\oint_{\alpha_{i}}\! f\d
u\Bigr)\Bigl(G-\oint_{\alpha_{i}}\!g\d u\Bigr)\biggr)\d z\biggr]\\
& = \sum_{i=1}^{r-1}\biggl[\oint_{\delta_{i}}\! pF\d
z\oint_{\alpha_{i}}\! g\d z + \oint_{\delta_{i}}\! pG\d
z\oint_{\alpha_{i}}\! f\d z\\ &\hskip 3cm
-\oint_{\delta_{i}}\! g\d
z\oint_{\alpha_{i}}\! pF\d z - \oint_{\delta_{i}}\! f\d
z\oint_{\alpha_{i}}\! pG\d z\biggr]\, .
\end{split}\ee
Using $\delta_{i}=\beta_{i}-\beta_{r}$ and
$\alpha=\sum_{i}\alpha_{i}$, we can rewrite the above formula as
\be\label{st1b}\begin{split}\mathscr I &=
\sum_{i=1}^{r}\biggl[\int_{\beta_{i}}\! pF\d z\oint_{\alpha_{i}}\! g\d
z + \int_{\beta_{i}}\! pG\d z\oint_{\alpha_{i}}\! f\d z
-\int_{\beta_{i}}\! g\d z\oint_{\alpha_{i}}\! pF\d z -
\int_{\beta_{i}}\! f\d z\oint_{\alpha_{i}}\! pG\d z\biggr]\\& \hskip
1cm -\int_{\beta_{r}}\! pF\d z\oint_{\alpha}\! g\d z -
\int_{\beta_{r}}\! pG\d z\oint_{\alpha}\! f\d z +\int_{\beta_{r}}\!
g\d z\oint_{\alpha}\! pF\d z + \int_{\beta_{r}}\! f\d
z\oint_{\alpha}\! pG\d z\, . \end{split}\ee
\begin{figure}
%\centerline{\includegraphics[width=3in]{fig3}}
\centerline{\epsfig{file=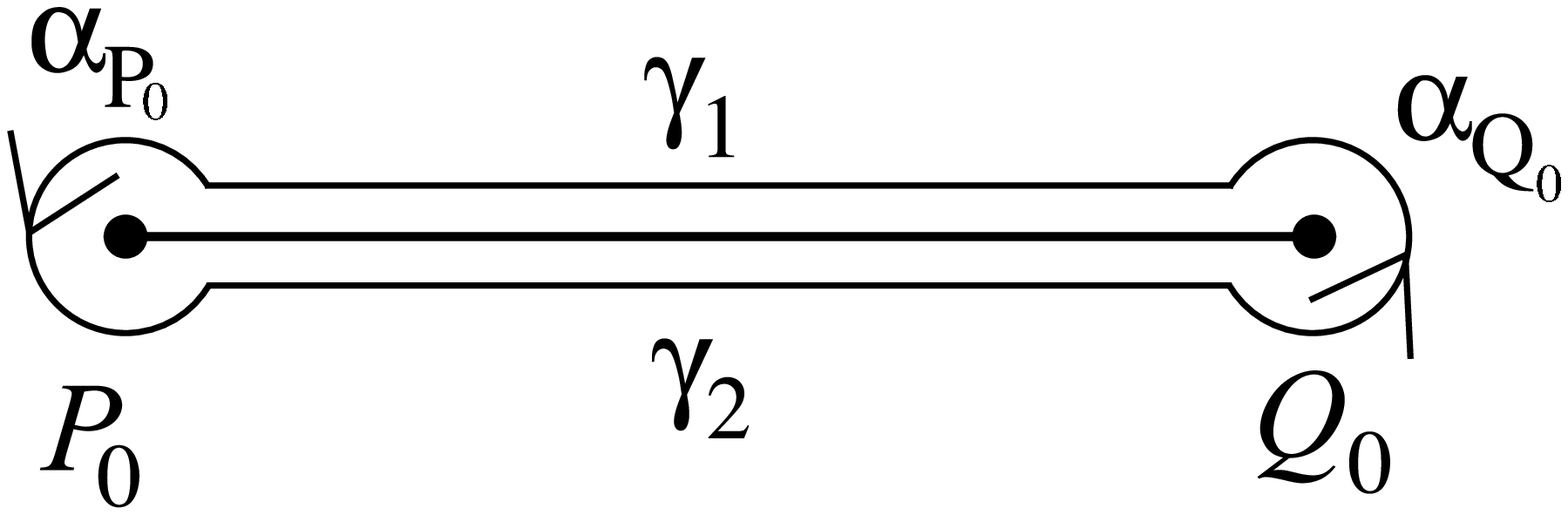,width=3in}}
\caption{Decomposition of the contour
$\gamma=\gamma_{1}+\alpha_{P_{0}}+\gamma_{2}+\alpha_{Q_{0}}$. The
contours $\alpha_{P_{0}}$ and $\alpha_{Q_{0}}$ are small circles
around the points at infinity. The contours $\gamma_{1}$ and
$\gamma_{2}$ go along the regularized contour $\beta_{r}$, joining the
points $z=\mu_{0}$ on the first and second sheets of the curve
$\mathcal C$.}\label{figcont}
\end{figure}

A second way to compute $\mathscr I$ is to deform the contour $\gamma$
so that it encircles the logarithmic branch cut from $P_{0}$ to
$Q_{0}$. We actually decompose
$\gamma=\gamma_{1}+\alpha_{P_{0}}+\gamma_{2}+\alpha_{Q_{0}}$ as
indicated in Figure \ref{figcont}. The discontinuity across the
logarithmic cut is given by the residues $f_{0}$ and $g_{0}$ of $f\d
z$ and $g\d z$ at $P_{0}$ in such a way that
\be\label{st2a} \int_{\gamma_{1}+\gamma_{2}}\!pFG\,\d z =
\int_{\beta_{r}}p\bigl(FG - (F-2i\pi f_{0})(G-2i\pi g_{0})\bigr)\d z\,
. \ee
Using the fact that the residues at infinity are given by minus the
integral of the corresponding forms over $\alpha=\sum_{i}\alpha_{i}$,
we thus obtain
\be\label{st2b}\mathscr I = -\oint_{\alpha}\!g\,\d
z\int_{\beta_{r}}\!pF\,\d z - \oint_{\alpha}\!f\,\d
z\int_{\beta_{r}}\!pG\,\d z
+\oint_{\alpha_{P_{0}}+\beta_{P_{0}}}\!pFG\,\d z\, . \ee
Putting \eqref{st1b} and \eqref{st2b} together, we find the
fundamental formula
\be\label{stfinal}\begin{split}
\sum_{i=1}^{r}&\biggl[\int_{\beta_{i}}\! pF\d z\oint_{\alpha_{i}}\!
g\,\d z + \int_{\beta_{i}}\! pG\d z\oint_{\alpha_{i}}\! f\d z
-\int_{\beta_{i}}\! g\,\d z\oint_{\alpha_{i}}\! pF\d z -
\int_{\beta_{i}}\! f\d z\oint_{\alpha_{i}}\! pG\d z\biggr]\\& \hskip
3cm +\int_{\beta_{r}}\! g\,\d z\oint_{\alpha}\! pF\d z +
\int_{\beta_{r}}\! f\d z\oint_{\alpha}\! pG\d z=
\oint_{\alpha_{P_{0}}+\beta_{P_{0}}}\!pFG\,\d z \, . \end{split}\ee
The right-hand side of \eqref{stfinal} is often conveniently rewritten
in terms of the analytic continuations $\hat F$ and $\hat G$ as
\be\label{Rint}\oint_{\alpha_{P_{0}}+\beta_{P_{0}}}\!pFG\,\d z =
-\oint_{\alpha}p(FG + \hat F\hat G)\d z\, .\ee

A relation more akin to the standard Riemann bilinear relations is
found when $F$ is single-valued on $\mathcal C$, i.e.\ when
\eqref{perf} is satisfied. Equation \eqref{stfinal} then reduces to
\be\label{stspecial}
\sum_{i=1}^{r}\biggl[\int_{\beta_{i}}\! pF\d z\oint_{\alpha_{i}}\!
g\,\d z -\int_{\beta_{i}}\! g\,\d z\oint_{\alpha_{i}}\! pF\d z \biggr]
+\int_{\beta_{r}}\! g\,\d z\oint_{\alpha}\! pF\d z =
-\oint_{\alpha}p(FG + \hat F\hat G)\d z\, .\ee
This latter formula can be deduced straightforwardly from the Riemann
bilinear relations found in textbooks.
\vfill\eject

\section{The solution in the rank one case}
\label{Ab}

In this Appendix, we discuss the explicit solution of the extended
theory in the special case corresponding to
\be\label{specialWV} W(z) = \frac{1}{2} mz^{2}\, ,\quad V(z) =
\la_{-1} + \la_{0} z + \frac{1}{2}\la_{1}z^{2}\, .\ee
In the usual case, for which $\la_{1}=\la_{0}=0$, the theory has $N$
confining vacua, the $N$-fold degeneracy corresponding to chiral
symmetry breaking. When $\la_{0}$ and $\la_{1}$ are turned on, we are
going to find generalizations of these vacua as well as new purely
quantum solutions that go to infinity in the classical limit.

The generating functions \eqref{Rdef} and \eqref{Sdef} are given by
\begin{align}\label{Rspesol} R(z) &=
\frac{N}{2}\biggl[\la_{0}+\la_{1}z + \frac{2+ 2s\la_{1}/m - \la_{0}z -
\la_{1}z^{2}}{\sqrt{z^{2} - 4s/m}}\biggr],\\\label{Sspesol} S(z)& =
\frac{Nm}{2}\Bigl[z-\sqrt{z^{2} - 4s/m}\Bigr] .\end{align}
The variable $s$, which coincides with the gluino condensate, is given
by the equation
\be\label{conspe} \int_{\beta_{1}}\!R\,\d z = NV(\mu_{0})-2N\ln\mu_{0}
+ 2i\pi k = N\int_{2\sqrt{s/m}}^{\mu_{0}} \frac{ \la_{1}z^{2}+\la_{0}z
-2s\la_{1}/m-2}{\sqrt{z^{2} - 4s/m}}\,\cdotp\ee
In terms of the instanton factor \eqref{qdef}, this is equivalent to
\be\label{qemspc} q = \Bigl(\frac{s}{m}\Bigr)^{N}e^{-N\la_{1}s/m}.\ee

Let us study the solutions to \eqref{qemspc} when $\la_{1}$ is small.
First, there are the usual $N$ solutions with small corrections,
\be\label{usuals} s = me^{2i\pi k/N}q^{1/N}\Bigl( 1 + \la_{1}e^{2i\pi
k/N}q^{1/N} + \mathcal O\bigl(\la_{1}^{2}\bigr)\Bigr),\quad 0\leq
k\leq N-1\, .\ee
More interestingly, there is also an infinite set of solutions that
have very large values of $s$, of the form
\be\label{notusuals} s\simeq
-\frac{m}{N\la_{1}}\Bigl(\ln\bigl(\la_{1}^{N}q\bigr) + 2i\pi k\Bigr)\,
,\quad k\in\mathbb Z\, . \ee
In terms of the Yang-Mills coupling constant $g_{\text{YM}}$ and theta
angle $\vartheta$, this takes the suggestive form
\be\label{notusuals2} s\simeq
\frac{m}{N\la_{1}}\Bigl(\frac{8\pi^{2}}{g_{\text{YM}}^{2}} -
i\vartheta-\ln\la_{1}^{N}\Bigr)\, . \ee
\end{document}